\documentclass[11pt]{article}
\usepackage[utf8]{inputenc}
\usepackage[T1]{fontenc}
\usepackage{indentfirst}
\usepackage{amsmath}
\usepackage{amssymb}
\usepackage{theorem}
\usepackage[margin=5em,font={small},position=bottom,labelfont=bf]{caption}
\usepackage{graphicx}
\usepackage{subfigure}
\usepackage{psfrag}

\newcommand{\eps}{\varepsilon}
\newcommand{\ket}[1]{{\left|#1\right\rangle}}
\newcommand{\bra}[1]{{\left\langle #1\right|}}


\renewcommand{\theequation}{\arabic{section}.\arabic{equation}}
\renewcommand{\thefigure}{\arabic{section}.\arabic{figure}}
\renewcommand{\thetable}{\arabic{section}.\arabic{table}}

\newcommand{\ssec}[1]{\section{#1}
\setcounter{equation}{0}
\setcounter{figure}{0}
\setcounter{table}{0}
}

\begin{document}

\flushright{ITP-Budapest Report No. 638}

\vspace{1cm}

\begin{center}
  {\LARGE L\"uscher's $\mu$-term and finite volume bootstrap principle for scattering states and form factors

}

\bigskip\bigskip

B. Pozsgay$^1$

\bigskip

{\it 
$^1$Institute for Theoretical Physics\\
E\"{o}tv\"os University, Budapest\\
H-1117 Budapest, P\'azm\'any P\'eter s\'et\'any 1/A
}
\bigskip

31th March 2008
\end{center}

\bigskip
\bigskip

\abstract{
We study the leading order finite size correction (L\"uscher's $\mu$-term) 
associated to moving one-particle states, arbitrary scattering states and finite
volume form factors in $1+1$ dimensional integrable models. 
Our method is based on the idea that the
$\mu$-term is intimately connected to the
inner structure of the particles, ie. their composition under the
bootstrap program. 
We 
use an appropriate analytic continuation of the Bethe-Yang
equations to quantize bound states in finite volume and obtain the
leading $\mu$-term (associated to symmetric particle fusions)
by calculating the deviations from the predictions of
the ordinary Bethe-Yang quantization.
Our results are compared to numerical data of the E$_8$ scattering theory
  obtained by truncated fermionic space approach.
As a by-product it is shown that the bound state quantization does not
only yield the correct $\mu$-term, but also provides the sum over a
subset of higher order corrections as well. 
}

\ssec{Introduction}

The knowledge of the properties of finite volume QFT is of  central
importance in at least two
ways. On one hand, numerical approaches to QFT
necessarily deal with a finite volume box and in order to interpret
the results correctly a reliable theoretical control of finite size corrections is needed.
On the other hand, working in finite volume is not
necessarily a disadvantage. On the contrary, the 
volume dependence of the spectrum can be exploited to obtain (infinite
volume) physical
quantities like the elastic
scattering phase shifts
\cite{luscher_scattering_states,luscher_szorodas_1+1} or resonance
widths \cite{luscher_bomlas,resonances}.  

Finite size mass correction  were first derived by L\"uscher  \cite{luscher_mass_corr}.
The  $1+1$ dimensional formulas relevant to integrable models together
with  a generalized F-term formula
for moving particles were obtained in \cite{klassen_melzer}.
Finite size corrections have recently become important in the context of AdS/CFT
correspondence as well. The
generalized $\mu$-term and F-term formulas for moving 
particles with arbitrary dispersion relation were derived in
\cite{janik_lukowski_1,janik_lukowski_2}. 

Besides the volume dependence of the spectrum itself, 
finite volume form factors (matrix elements of
local operators) 
also represent a central object in finite volume QFT. Apart from the
obvious relevance to lattice QFT they are also important
in $1+1$
dimensional models, where they can be
used to construct a systematic low-temperature expansion of
correlation functions at finite temperature \cite{fftcsa2}. 
The connection
to infinite volume form factors is given by a simple (though non-trivial)
proportionality factor
\cite{lellouch_luscher,k_to_pipi,fftcsa1} which is exact to all orders
in the inverse of the volume. There are
however finite size corrections that decay exponentially with the volume and they
play a crucial role whenever the numerical simulation is limited to
small volumes. Moreover, they can produce huge deviations even in
relatively large volumes provided that the exponent is small. 
This work was partly motivated by such an example which can be found
in  \cite{fftcsa1} (section 4.1.2).

In this work we present a method to obtain the leading 
$\mu$-term associated to arbitrary multi-particle energy levels and
finite volume form factors in $1+1$
dimensional integrable models. 
Our approach is based on the 
idea that the $\mu$-term is associated to the ,,inner structure'' of
the particles, ie. their composition under the bootstrap program. It
is supposed that the leading $\mu$-term is caused by a symmetric
particle fusion $A_aA_a\to A_c$.
 The results can also be applied in nonintegrable models
for states below the first inelastic threshold.

The outline of the paper is as follows.

We begin our analysis in sec. 2 by giving a new interpretation of L\"uscher's
$\mu$-term and extending it to describe moving particles. 
The scaling Ising model serves as a testing ground for our
calculations: the analytic predictions are compared to numerical data
obtained by the Truncated Conformal Space Approach (TCSA) developed by
Yurov and Zamolodchikov \cite{yurovzamTCSA}. 
The extension of our results to arbitrary multi-particle scattering
states is presented in section 3.

Section 4 deals with the $\mu$-term of finite volume form factors and
section 5 is devoted to the conclusions.

\ssec{One-particle states}

\subsection{Bound-states in finite volume}

Let us consider an integrable QFT with diagonal scattering. The spectrum consists
of particles $A_i$, $i=1,\dots,N$, with masses $m_i$ which are assumed
to be strictly non-degenerate. Asymptotic states are denoted by
\begin{equation*}
  \ket{\theta_1,\theta_2,\dots,\theta_n}_{i_1i_2\dots i_n},
\end{equation*}
where the indices $i_1\dots i_n$ denote the particle species.
Multi-particle scattering processes are described by the products of
two-particle phase shifts $S_{ij}(\theta_{ij})$ where $\theta_{ij}$ is the
relative rapidity of the incoming particles $A_i$ and $A_j$.

\begin{figure}
   \centering
\psfrag{Aa}{$A_c$}
\psfrag{Ab}{$A_a$}
\psfrag{Ac}{$A_b$}
\psfrag{a1}{$A_1$}
\psfrag{a2}{$A_2$}
\psfrag{a5}{$A_5$}
\psfrag{sim}{$\sim$}
\psfrag{ubca}{$\bar{u}_{ac}^b$}
\psfrag{ucba}{$\bar{u}_{bc}^a$}
\psfrag{mu}{$\mu_{ab}^c$}
\psfrag{ma}{$m_c$}
\psfrag{mb}{$m_a$}
\psfrag{mc}{$m_b$}
\footnotesize
\subfigure[Fusion angles]{\includegraphics[bb=-80 -20 290 277,scale=0.4]{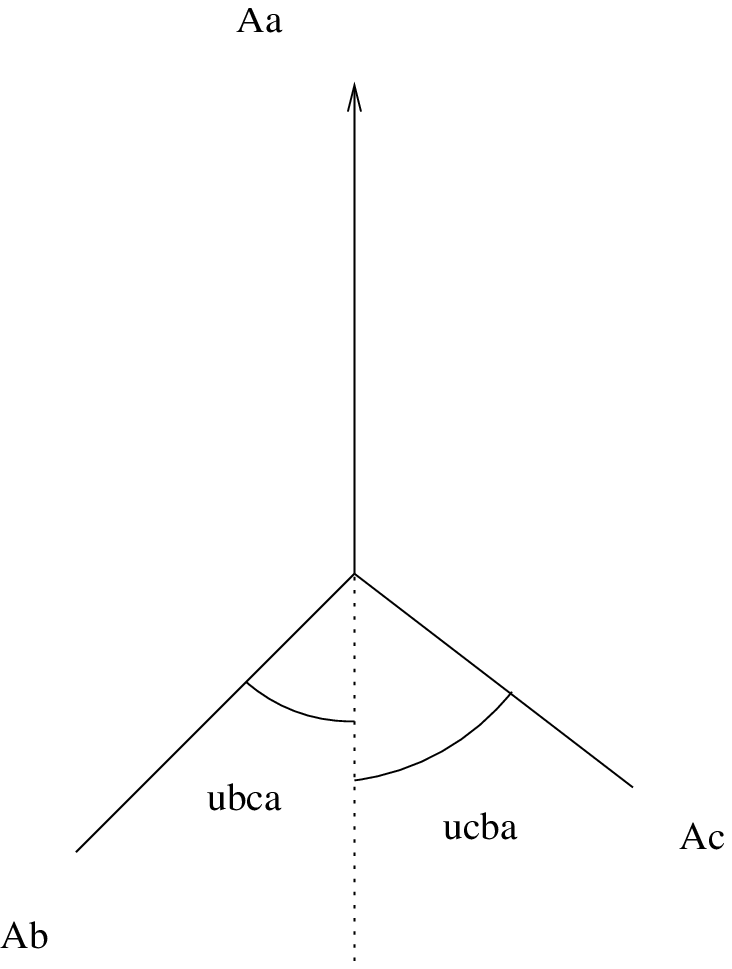}}
\subfigure[The mass triangle]{\includegraphics[bb=-110 -50 227 145,scale=0.4]{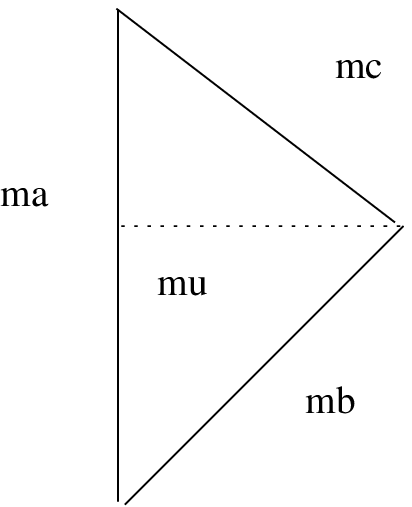}}
\subfigure[A triple bound state (see 2.2.3. for explanation)]{\includegraphics[bb=-80 -30 386 279,scale=0.30]{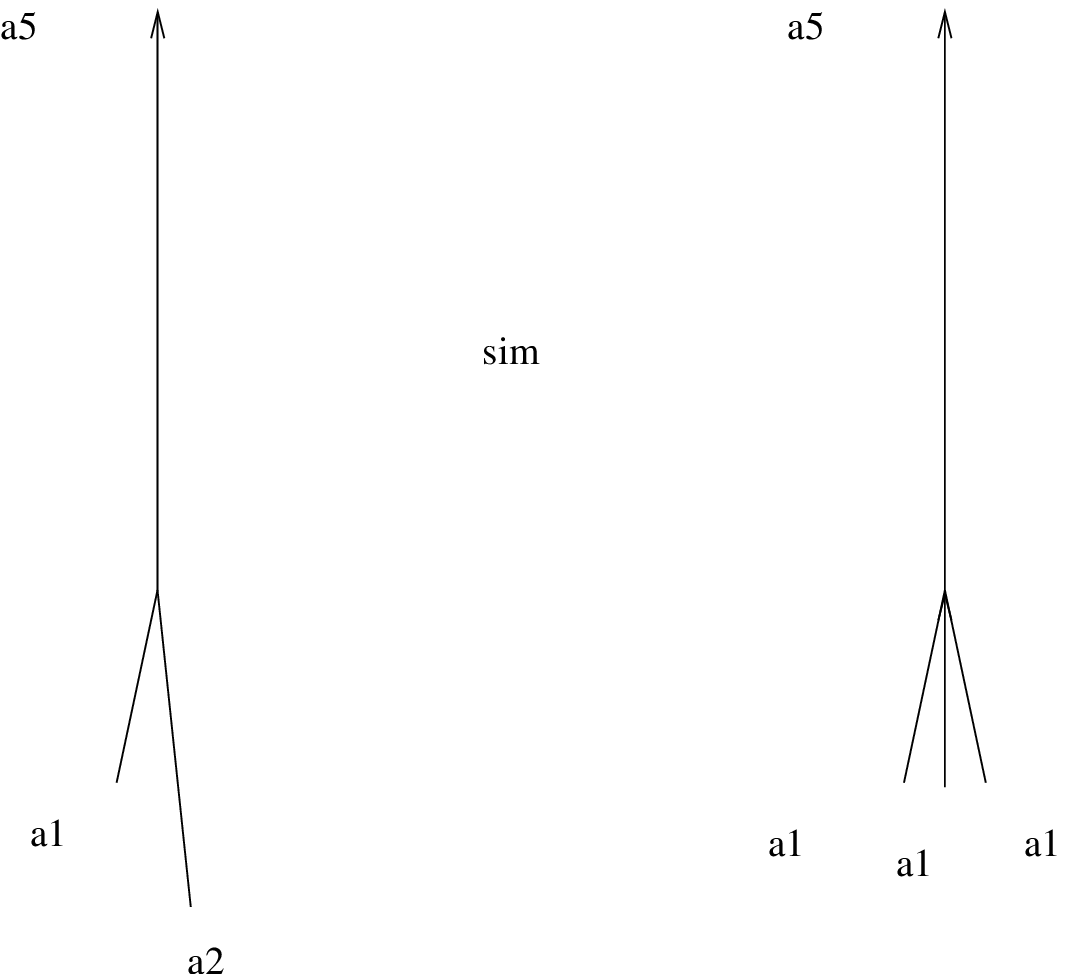}}
\caption{Pictorial representation of particle fusions.\label{fusion}}
 \end{figure}

The explicit formulas for the leading finite size mass corrections in a periodic box of
volume $L$ read  \cite{klassen_melzer}
\begin{eqnarray}
\label{mass_mu_term}
\Delta m_a^{(\mu)}&=&
-\mathop{\sum\nolimits'}_{b,c}\theta(m_a^2-|m_b^2-m_c^2|)\mu_{ab}^c
\left(\Gamma_{ab}^c\right)^2 e^{-\mu_{ab}^cL}\\
\label{mass_F_term}
\Delta m_a^{(F)}&=&-\mathop{\sum\nolimits'}_b \mathcal{P} \int_{-\infty}^\infty
\frac{d\theta}{2\pi}e^{-m_bL\cosh(\theta)}m_b \cosh(\theta)
\left(S_{ab}(\theta+i\pi/2)-1\right)
\end{eqnarray}
$\mu_{ab}^c$ is the altitude of the mass triangle with base
$m_c$ (see figure \ref{fusion}) and $\left(\Gamma_{ab}^c\right)^2$ is the
 residue of $S_{ab}(\theta)$ corresponding to the formation of the
 bound state.

Here we determine the leading $\mu$-term associated to a moving one-particle state $A_c$.
 Based on the description of 
mass corrections it is expected that this contribution is associated 
to the fusion $A_aA_b\to A_c$ with the smallest $\mu_{ab}^c$.
We assume that $a=b$, ie. the fusion in question is a symmetric
one.
This happens to be true for the lightest particle in models with the ''$\Phi^3$-property'' 
and for other low lying states in most known models. At the end of
this section we comment on the possible extension to nonsymmetric fusions.

The bootstrap principle for a symmetric fusion consists of the identification
\begin{equation}
\label{infi_bootstrap}
  \ket{\theta}_c \sim \ket{\theta+i\bar{u}_{ac}^a,\theta-i\bar{u}_{ac}^a}_{aa}
\end{equation}
resulting in
\begin{equation*}
  m_c=2m_a\cos(\bar{u}_{ac}^a) \quad\quad \left(\mu_{aa}^c\right)^2=m_a^2-\frac{m_c^2}{4}
\end{equation*}
Smallness of $\mu_{aa}^c$ means that $m_c$ is close to $2m_a$, in
other words the
binding energy is small. 

For a moment let us lay aside the framework of QFT
and consider quantum mechanics with an
attractive potential. Bound states are described by 
wave functions 
\begin{equation*}
\Psi(x_1,x_2)=e^{iP(x_1+x_2)}\psi(x_1-x_2)
\end{equation*}
where $P$ is the total momentum and $\psi(x)$ is the appropriate
solution of the Schr\"odinger equation in the relative coordinate. It is localized around $x=0$ and
shows exponential decay at infinity. Except for the
region $x_1\approx x_2$, the  wave function can be approximated
with a product of plane waves with imaginary momenta $p_{1,2}=P\pm ik$. The interaction
results in the quantization of the allowed values of $k$.

The theory in finite volume is described along the same lines. There
are however two differences:
\begin{itemize}
\item The total momentum gets quantized.
\item $\psi(x)$ (and therefore $k$) obtains finite volume corrections.
\end{itemize}

This picture also applies to relativistic integrable
theories. We consider $A_c$ as a simple quantum mechanical
bound state of two elementary particles and  use the infinite volume
scattering data to describe the interaction between the constituents.
To develop these ideas, let us consider the spectrum of the theory
defined on a circle with circumference $L$. We state the identification 
\begin{equation}
  \label{finiteL_azonositas}
\ket{A_c(\theta)}_L\sim  \ket{A_a(\theta_1)A_a(\theta_2)}_L
\end{equation}
where the $\theta_{1,2}$ are complex to describe a
bound-state; this idea also appeared in \cite{takacs_watts,bpt1}. 
Relation \eqref{finiteL_azonositas} can be regarded
as  the finite volume realization of \eqref{infi_bootstrap}. 
 The total energy and momentum of the bound state have to be purely real,
constraining the rapidities to take the form
\begin{equation}
\label{finiteL_rapidities}
  \theta_1=\theta+iu,\quad\quad
  \theta_2=\theta-iu
\end{equation}
where the dependence on $L$ is suppressed. Energy and momentum
are calculated as
\begin{equation}
  E=2m_a\cos(u)\cosh(\theta)\quad\quad p=2m_a\cos(u)\sinh(\theta)
\end{equation}

The momenta of
two-particle states in finite volume are quantized by a relation
involving the scattering phase shift
\cite{luscher_scattering_states}. This procedure can be extended
in $1+1$
dimensional integrable models 
to arbitrary multi-particle scattering states. The quantization
condition for an $n$-particle state is given by the Bethe-Yang equations 
\begin{equation*}
  e^{ip_jL}\mathop{\prod_{k=1}^n}_{k\ne j}
S_{i_ji_k}(\theta_j-\theta_k)=1,\quad\quad
j=1\dots n
\end{equation*}
To
quantize the bound state in finite volume, an appropriate analytic
continuation of the above equations with $n=2$ can be applied.
This procedure is justified
by the same reasoning that leads to original Bethe-Yang equations: one
assumes plane waves (with imaginary momenta) except for
the localized interaction, which is described by the S-matrix of the
infinite volume theory.
Inserting \eqref{finiteL_rapidities} and separating the real and imaginary
parts 
\begin{eqnarray}
\label{kepzetes_Bethe_Yang_1}
e^{im_a \cos(u)\sinh(\theta)L} e^{-m_a\sin(u)\cosh(\theta)L}
S_{aa}(2iu)&=&1\\
\label{kepzetes_Bethe_Yang_2}
e^{im_a \cos(u)\sinh(\theta)L} e^{m_a\sin(u)\cosh(\theta)L}
S_{aa}(-2iu)&=&1
\end{eqnarray}
Multiplying the two equations and making use of $S(2iu)=S(-2iu)^{-1}$ one arrives at
\begin{equation}
\label{kvantalasi_1}
  e^{2im_a \cos(u)\sinh(\theta)L}=1\quad \textrm{or}\quad  2m_a \cos(u)\sinh(\theta)=\frac{2\pi I}{L}
\end{equation}
which is the quantization condition
for the total momentum. $I$ is to be identified with the momentum quantum number
of $A_c$.  The quantization condition for $u$ is found by eliminating
$\theta$ from \eqref{kepzetes_Bethe_Yang_1}:
\begin{equation}
  \label{u_egyenlete}
  e^{-m_aL\sin(u)\sqrt{1+\left(\frac{\pi I}{m_aL\cos(u)}\right)^2}}
S_{aa}(2iu)=(-1)^I
\end{equation}
The exponential factor forces $u$ to be close to the pole of
the S-matrix associated to the formation of the bound-state. For the
case at hand it reads
\begin{equation}
\label{S-matrix_polus}
S_{aa}(\theta\sim iu_{aa}^c)\sim \frac{i\left(\Gamma_{aa}^c\right)^2}{\theta-iu_{aa}^c}
\end{equation}
with $u_{aa}^c=2\bar{u}_{ac}^a$.
Note the appearance of $(-1)^I$ on the rhs. of \eqref{u_egyenlete}, which is a natural
consequence of the quantization of the total momentum. This sign
determines the direction from which the pole is approached.

The exact solution of \eqref{u_egyenlete} can be developed into a power series in
$e^{-\mu_{aa}^c L}$, where the first term is found by replacing $u$
with $\bar{u}_{ac}^a$ in the exponent:
\begin{equation}
\label{uminusubar}
  u-\bar{u}_{ac}^a=(-1)^I \frac{1}{2}\left(\Gamma_{aa}^c\right)^2
  e^{-\mu_{aa}^cL \sqrt{1+\left(\frac{2\pi I}{m_cL}\right)^2}}+
O(e^{-2\mu_{aa}^cL})
\end{equation}
First order corrections to the energy
are readily evaluated to give
\begin{eqnarray}
\label{elso_korrekcio}
  E&=&E_0- (-1)^I  \left(\Gamma_{aa}^c\right)^2
\frac{\mu_{aa}^c m_c}{E_0} e^{-\frac{\mu_{aa}^cE_0}{m_c} L }+O(e^{-2\mu_{aa}^cL})
\end{eqnarray}
where $E_0$ is the ordinary  one-particle energy
\begin{equation*}
  E_0=\sqrt{m_c^2+\left(\frac{2\pi I}{L}\right)^2}
\end{equation*}
In the case of zero momentum the former result simplifies to 
the leading term in \eqref{mass_mu_term}. 
For large volumes we
recover
\begin{equation*}
u\to \bar{u}_{ac}^a\quad\quad  \theta\to \textrm{arsh} \frac{2\pi I}{m_cL}
\end{equation*}

Having established the quantization procedure we now turn to the
question of momentum quantum numbers inside 
the bound state. 
For the phase shift let us adopt the convention introduced in \cite{fftcsa1}
\begin{equation*}
  S_{ab}(\theta)=S_{ab}(0)e^{i\delta_{ab}(\theta)}
\end{equation*}
where $\delta_{ab}(\theta)$ is defined to be continuous on the real
line and is antisymmetric by unitarity and real analyticity.
$\delta_{ab}(\theta)$ can be extended unambiguously to the imaginary
axis 
by analytic continuation apart from the choice of the logarithmic branch.
For a generic rapidity one has
\begin{equation*}
  \delta_{ab}(\theta^*)=\delta_{ab}(\theta)^*\quad\quad \delta_{ab}(-\theta)=-\delta_{ab}(\theta)
\end{equation*}
Note that $\delta_{ab}(iu)$ is purely imaginary.

With this choice of the phase shift the Bethe-Yang equations in their logarithmic form 
\begin{eqnarray*}
  l \sinh(\theta+iu) +\delta_{11}(2iu)&=&2\pi I_1\\
  l \sinh(\theta-iu) +\delta_{11}(-2iu)&=&2\pi I_2
\end{eqnarray*}
imply $I_1=I_2$. Quantization of the total momentum on the other hand requires
$I_1=I_2=I/2$. Note that different conventions for $\delta_{ab}$ would result in a less
transparent rule for dividing $I$ among the two constituents. The only
disadvantage of our choice is the appearance of the unphysical half-integer quantum
numbers.  

Let us denote a multi-particle state in finite
volume as
\begin{equation*}
  \ket{\{I_1,\dots I_n\}}_{i_1i_2\dotsi_n,L}
\end{equation*}
where the quantum numbers $I_i$ serve as an input to the Bethe-Yang
equations. The bound-state quantization can be written in
short-hand notation as
\begin{equation*}
  \ket{\{I\}}_{c,L}\sim\ket{\{I/2,I/2\}}_{aa,L}
\end{equation*}

The example of mass corrections suggests that in order to obtain the
total $\mu$-term it is necessary to include  
a sum in  \eqref{elso_korrekcio} over the different fusions leading to
$A_c$. 
However, the case of nonsymmetric fusions requires special care.
Here we outline the
difficulties of the nonsymmetric bound state quantization.

A straightforward
application of the Bethe-Yang equations to an
$\ket{A_a(\theta+i\bar{u}_{ac}^b)A_b(\theta-i\bar{u}_{bc}^a)}_L$  bound state
yields a
nontrivial phase factor $e^{2I\pi\frac{m_a}{m_a+m_b}}$ instead of
$(-1)^I$ in  \eqref{u_egyenlete}. This in turn implies that the
rapidity difference of the constituents can not be purely
imaginary. The real parts of the rapidities thus get different finite
size corrections and the total energy of the bound state becomes
complex. A possible
solution would be to take twice the real part of the energy
correction, corresponding to the sum of the contributions coming from
the $A_aA_b\to A_c$ and $A_bA_a\to A_c$ fusions. This is however only a
 guess and a more systematic
treatment is needed. In fact, the infinite volume bootstrap principle
suggests that the states 
\begin{equation*}
\ket{A_a(\theta-i\bar{u}_{ac}^b)A_b(\theta+i\bar{u}_{bc}^a)}_L \quad\textrm{and}\quad
\ket{A_a(\theta+i\bar{u}_{ac}^b)A_b(\theta-i\bar{u}_{bc}^a)}_L
\end{equation*}
should be handled on an equal footing. A possible way to accomplish
this would be to develop a multi-channel Bethe-Yang quantization
scheme. However, this is beyond the scope of the present work.

As a conclusion of this section \eqref{uminusubar} is compared to the lowest order results of the
TBA approach. The general discussion of excited states TBA equations
in diagonal scattering theories is not available. For simplicity we
restrict ourselves to the Lee-Yang model which was considered in the
original paper \cite{excited_TBA}. In this nonunitary model there is
only one
particle and the scattering is described by
\begin{equation*}
  S(\theta)=\frac{\sinh(\theta)+i\sin(\pi/3)}{\sinh(\theta)-i\sin(\pi/3)}
\end{equation*}
The exact TBA equations for moving one-particle states read
\begin{eqnarray}
\label{TBA_E}
  E&=&-im(\sinh\theta_0-\sinh\bar{\theta}_0)-
\int_{-\infty}^\infty\frac{d\theta}{2\pi} m\cosh(\theta) L(\theta)\\
\eps(\theta)&=&mR \cosh\theta
+\log\frac{S(\theta-\theta_0)}{S(\theta-\bar{\theta}_0)} 
-(\varphi\star L) (\theta)
\end{eqnarray}
where
\begin{equation*}
  L(\theta)=\log(1+e^{-\eps(\theta)})\quad\textrm{and}\quad 
(f\star g)(\theta)= \int_{-\infty}^\infty\frac{d\theta'}{2\pi}
f(\theta-\theta') g(\theta')
\end{equation*}
Here the volume is denoted by $R$ to avoid confusion with $L(\theta)$.
The complex rapidity $\theta_0$ satisfies the consistency equation 
\begin{equation}
\label{TBA_consist}
  \eps(\theta_0)=mR \cosh\theta_0+i\pi
-\log(S(2i\textrm{Im}\theta_0))-(\varphi\star L) (\theta)=i(2n+1)\pi
\end{equation}
The convolution term in \eqref{TBA_consist} can be neglected and one
obtains $\textrm{Im}\theta_0=\pi/6+\delta$ where 
$\delta$ is exponentially small. To zeroth order one also has
\begin{equation*}
mR\cosh(\textrm{Re}  \theta_0)=(4n+(1-\textrm{sign}\delta))\pi
\end{equation*}
which is the ordinary one-particle quantization condition with
$I=2n+\frac{1}{2}(1-\textrm{sign}\delta)$. 
Neglecting the contribution
of the integral in \eqref{TBA_E} and substituting $\theta_0=\theta+iu$
one has $E=2m\sin(u)\cosh(\theta)$. This is exactly the energy of an $AA$
bound state with the imaginary rapidities $\theta\pm iu$. Separating
the real and imaginary parts of \eqref{TBA_consist} and still neglecting
the convolution term (which is responsible for the F-term) one obtains equations \eqref{u_egyenlete} and
\eqref{kvantalasi_1}, thus proving the consistency of the two approaches.

It seems plausible that  by
exactly solving the bound state quantization condition \eqref{u_egyenlete} one obtains all higher order
corrections that go as $e^{-n\mu_{aa}^cL}$ with $n\in \mathbb{N}$.  In the
next subsection we present numerical evidence to support this claim.

\subsection{Numerical analysis}

We investigate the famous E$_8$  scattering
theory \cite{zamE8}, which is the relativistic integrable field theory associated to
the scaling limit of the Ising model in the presence of a magnetic
field. 
The infinite volume spectrum of the model consists of 8 particles.
The first three particles lie below the two-particle threshold and they all show up as $A_1A_1$
bound states. These fusions are responsible for the leading $\mu$-term.
The corresponding parameters (in units of $m_1$)
are listed in the table below. The exponent of the next-to-leading
correction (the error exponent) is denoted by $\mu'$.

\begin{center}
\begin{tabular}{|c|ccc|cc|}
\hline
a& $m_a$ &  $\mu_{11}^a$ & $\left(\Gamma_{11}^a\right)^2$ & $\mu'$ & \\
\hline
1 & 1 & 0.86603 & 205.14 & $1$ & ($m_1$)\\

2 & 1.6180 & 0.58779 & 120.80 & 0.95106 & ($\mu_{12}^2$) \\

 3&  1.9890 & 0.10453& 1.0819 &  0.20906 &  ($2\mu_{11}^3$) \\
\hline
\end{tabular}
\end{center}

In \cite{klassen_melzer} Klassen and Melzer performed the numerical
analysis of mass corrections. The analytic predictions were compared to TCSA data
and to transfer matrix results. They
observed the expected behaviour of mass corrections of $A_1$ and
$A_2$; in the former case they were also able to verify the F-term. On
the other hand, the precision of their TCSA data was not sufficient to
reach volumes where the $\mu$-term for $A_3$ could have been tested. 
This limitation is a natural consequence of the unusually small
exponent $\mu_{11}^3$: the next-to-leading contribution is of order
$e^{-2\mu_{11}^3L}$, still very slowly decaying.

Here we employ
the TFCSA (Truncated Fermionic Conformal Space Approach, see \cite{yurovzam_fermionic_TFCSA})
routines that were successfully used in \cite{fftcsa2,fftcsa1}. 
Calculations are performed for $I=0,1,2,3$ at different values of
the volume. One-particle states of $A_1$, $A_2$ and $A_3$ are easily
identified: they are the lowest lying levels in the spectrum, except
for $I=0$ where the lowest state is the vacuum. 
We use the dimensionless quantities $l=m_1L$ and $e=E/m_1$.  

The results are extrapolated from $e_{cut}=20..30$ to
$e_{cut}=\infty$ using the procedure developed in
\cite{takacs_extrapolation}. Our experience shows
that this extrapolation technique reduces the numerical
errors by an order of magnitude. However, our attempts to develop an adequate method to
estimate these errors have failed. 
Several examples were encountered where the actual numerical deviation
was either underestimated or overestimated, no matter which 
estimate was used.
We therefore resign from quantitatively monitoring the TCSA errors and
constrain ourselves to a range
of the volume parameter where it is safe to neglect truncation effects.

\subsubsection{$A_3$}

We begin our analysis with the most interesting case of $A_3$. At each
value of $l$ and $I$ the following procedure is performed.
\begin{itemize}
\item The energy correction is calculated according to
  \eqref{elso_korrekcio}
\item The quantization condition \eqref{u_egyenlete} is solved for
  $u$ and the energy correction is calculated by
  \begin{equation*}
 \Delta e=2\cosh(\theta)\cos(u)-e_0
\end{equation*}
where $\theta$ is determined by the total momentum quantization
\eqref{kvantalasi_1} and $e_0$ is the  ordinary one-particle energy.
\item The exact correction is calculated numerically by $\Delta e=e^{TCSA}-e_0$.
\end{itemize}

The choice for the range of the volumes is limited in two ways. On one hand,
$l$ has to be sufficiently large in order to reduce the contribution of the
F terms and other higher order
finite size corrections. On the other hand, numerical errors grow with the volume and
eventually become
comparable with the finite size corrections, resulting in an upper bound
on $l$.  The window $l=30..40$ is suitable for
our purposes.

\begin{figure}
% a0_3_elso_elemzes.plt
  \centering
\subfigure[$I=0$]{  \includegraphics[scale=0.5,angle=-90]{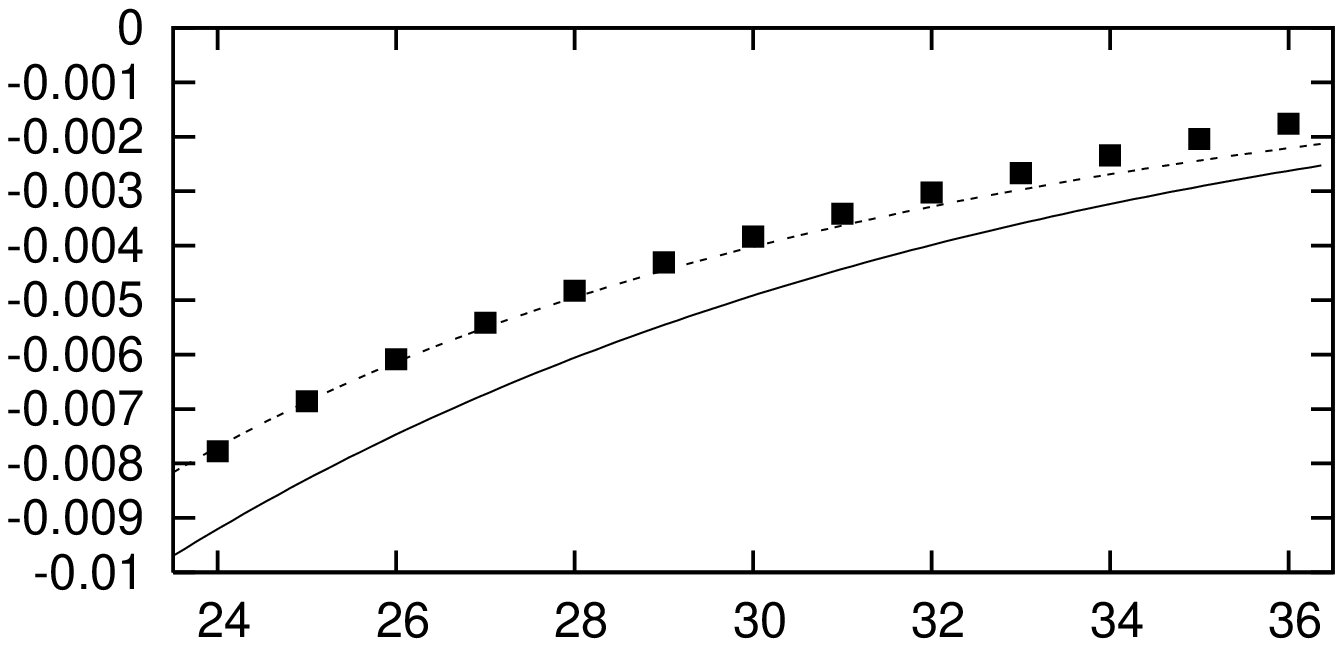}}
\subfigure[$I=1$]{ \includegraphics[scale=0.5,angle=-90]{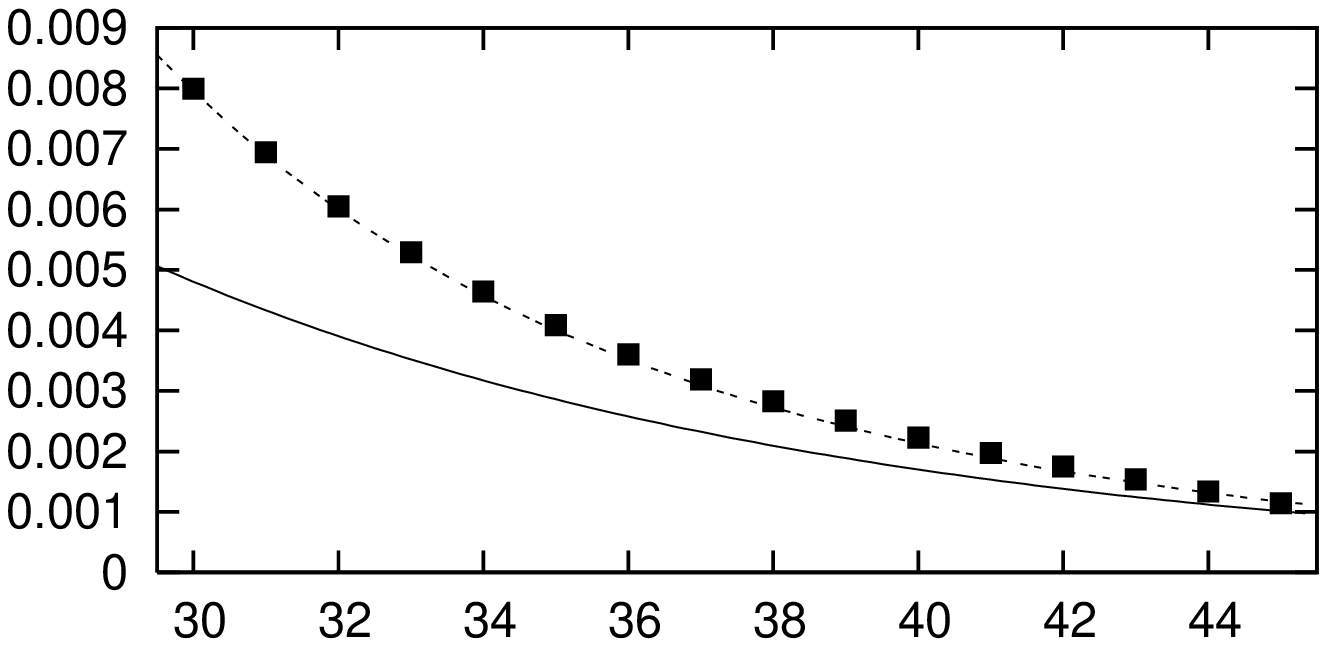}}

\subfigure[$I=2$]{ \includegraphics[scale=0.5,angle=-90]{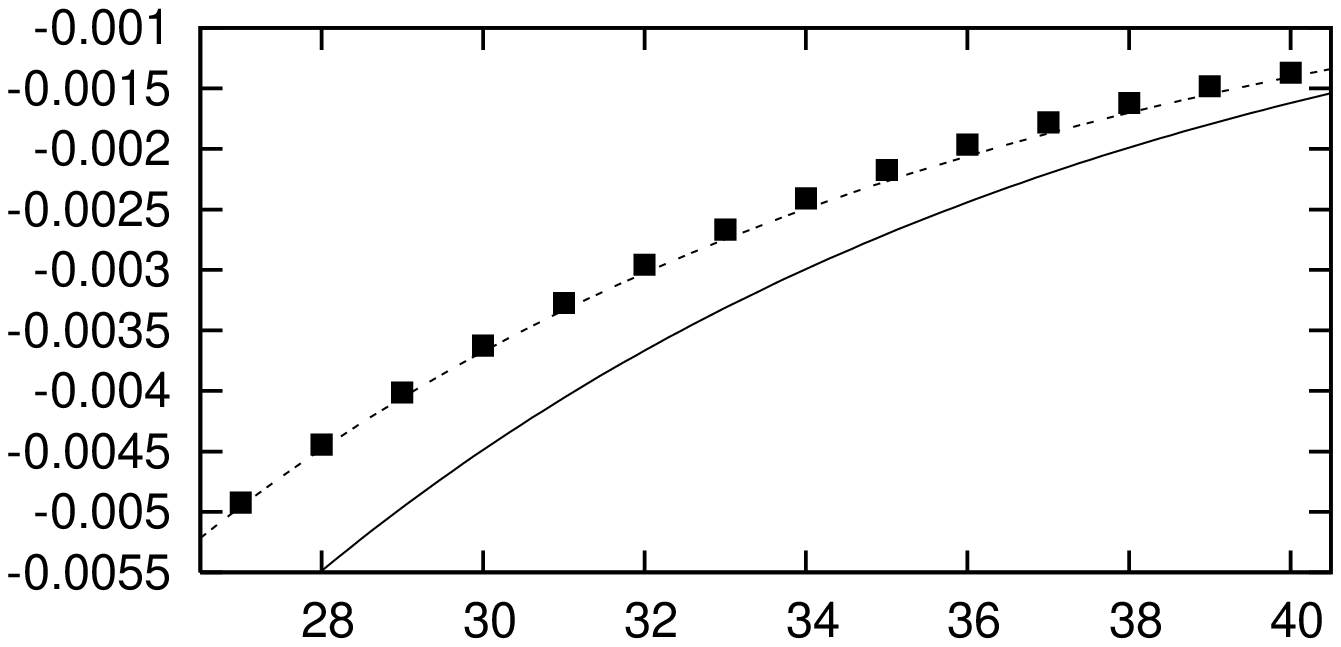}}
\subfigure[$I=3$]{\includegraphics[scale=0.5,angle=-90]{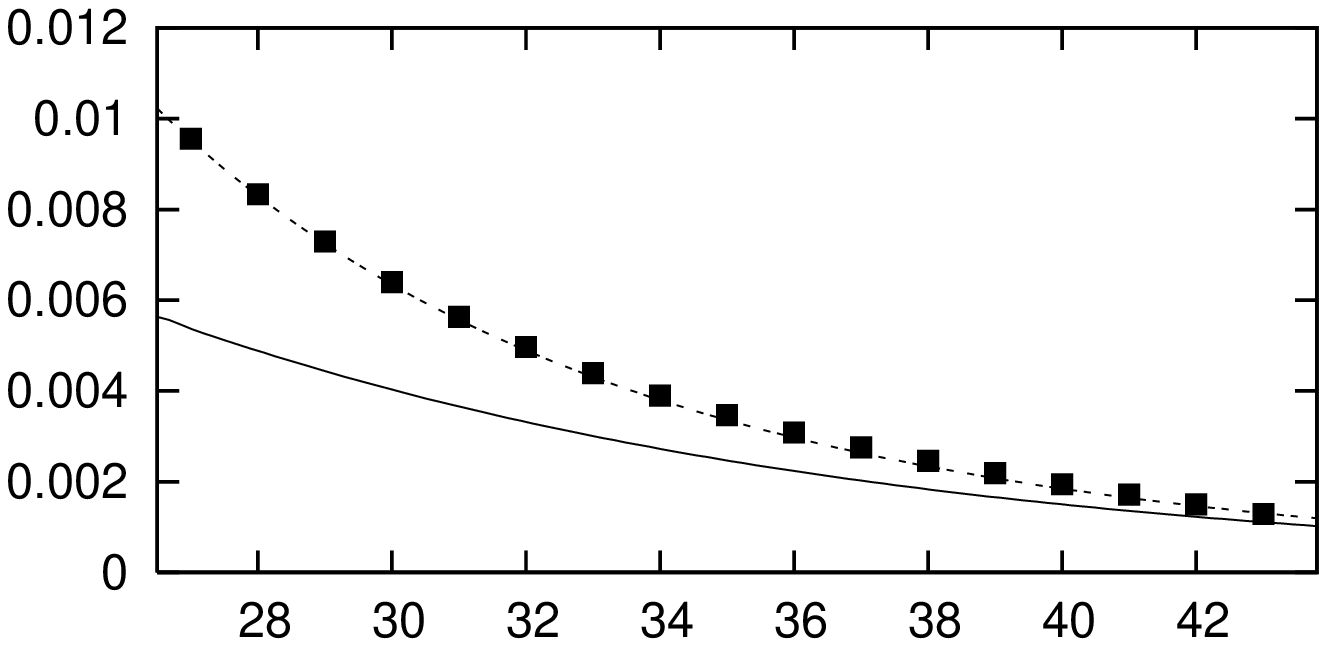}}
\caption{
Finite size corrections to $A_3$ one-particle levels (in sectors
$I=0\dots 3$) as a function of
the volume. The TCSA data
are plotted against theoretical predictions of the single
$\mu$-term associated to the $A_1A_1\to A_3$ fusion (solid curve) and
the exact solution of the bound-state quantization (dotted curve). 
\label{fig:A3_nagy_terfogat}}
\end{figure}

The results are shown in figure \ref{fig:A3_nagy_terfogat}. It is
clear that the $\mu$-term yields the correct prediction in the
$L\to\infty$ limit. However,  higher order terms cause a
significant deviation for $l<40$, which is in turn accurately desribed
by the bound state prediction.
The sign of the correction depends on
the parity of $I$ as predicted by \eqref{elso_korrekcio}.

Based on the success of this first numerical test we also explored the
region $l<30$.
Inspecting the behaviour of $u$ as a function of $l$ reveals an interesting
phenomenon. It is obvious from \eqref{uminusubar} that $u(l)$ is
monotonously increasing if $I$ is odd, with the infinite volume
limit fixed to $\bar{u}_{ac}^a$.  However, the complex
conjugate pair $\theta_{1,2}$ approaches the real axis as $l$ is
decreased and they collide at a critical volume $l=l_c$. For $l<l_c$ 
they separate again but stay on the real line, providing a unique
solution with two distinct purely real rapidities. The same behaviour
was also observed in \cite{takacs_watts,bpt1}.

The interpretation of this phenomenon is evident: if the volume is comparable to
the characteristic size of the bound-state, there is enough energy in
the system for the constituents to become unbound. 
Therefore the $A_3$ one-particle level becomes an $A_1A_1$
scattering state for $l<l_c$. We call this phenomenon the
``dissociation of the bound state''. The same result was
obtained also in the boundary sine-Gordon model by a semiclassical analysis \cite{bpt_semiclass}. 

The value of $l_c$ can be found
by exploiting the fact that the Jacobian of the Bethe-Yang equations (viewed as a mapping from
$(\theta_1,\theta_2)$ to $(I_1,I_2)$) vanishes at the critical
point. A straightforward calculation yields
\begin{equation*}
  l_c=\sqrt{4\varphi_{11}(0)^2-I^2\pi^2}
\end{equation*}
where $\varphi_{11}(\theta)=\delta_{11}'(\theta)$. 
The numerical values for the case at hand
are 
\begin{equation*}
  l_c=27.887\quad (I=1) \quad\textrm{ and }  \quad
l_c=26.434 \quad (I=3)
\end{equation*}

\begin{figure}
  \centering
% tomeg3.plt
\Large
\psfrag{spin0}{$I=0$}
\psfrag{spin1}{$I=1$}
\psfrag{spin2}{$I=2$}
\psfrag{spin3}{$I=3$}
\psfrag{l}{$l$}
\psfrag{e}{$e$}
\psfrag{igazi}{$m_3$}
\psfrag{magy1}{$A_1A_1\to A_3$}
\psfrag{magy2}{$(I=3)\quad (I=1)$}
\includegraphics[angle=-90,scale=0.5]{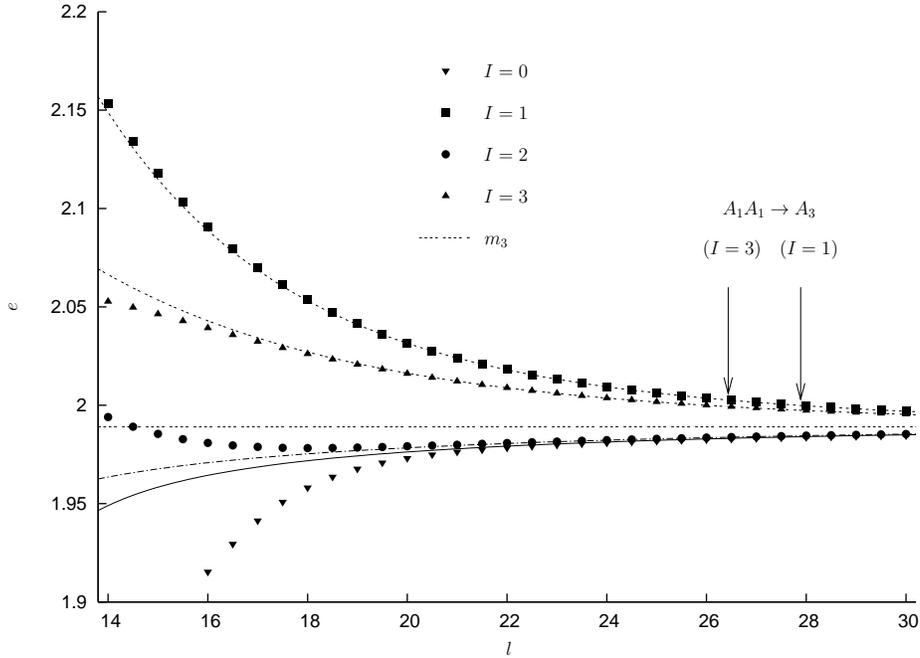}  
\caption{
$A_3$ one-particle levels in sectors $I=0\dots 3$  as a function of
the volume. Dots represent TCSA data, while the lines show the corresponding
prediction of the $A_1A_1$ bound state quantization. In sectors $I=1$ and
$I=3$ the bound state dissociates at $l_c$ and for $l<l_c$
 a conventional $A_1A_1$ scattering state replaces $A_3$ in the spectrum. 
The values of $l_c$ are shown by the two arrows.
\label{A3_full}}
\end{figure}

We are now in the position to complete the numerical analysis.
The Bethe-equations are solved at each value of $l$, 
providing two distinct real rapidities for $l<l_c$ (with $I$ being odd),
and a complex conjugate pair otherwise. The energy is calculated in
either case as
\begin{equation*}
  e=\cosh(\theta_1)+\cosh(\theta_2)
\end{equation*}
which is compared to TCSA data. The results are exhibited in figure 
\ref{A3_full}.

The agreement for the upper two curves ($I=1$ and $I=3$) is not as suprising as it may
seem because what one sees here are conventional $A_1A_1$ scattering
states. The Bethe-Yang equation determining their energy is exact up to $O(e^{-\mu'L})$ where 
$\mu'=\mu_{11}^1$
is the smallest exponent that occurs in the sequence of finite volume
corrections of $A_1$.
On the other hand, the energy levels are analytic functions of L, which
leads to the conclusion that the prediction of \eqref{u_egyenlete} is correct up to
$O(e^{-\mu_{11}^1L})$ even for $L>L_c$.  Comparing the numerical
values one finds $\mu_{11}^1>8\mu_{11}^3$. We conclude
 that the bound state picture  indeed accounts for
finite volume corrections up to the first few orders in
$e^{-\mu_{aa}^cL}$ (the first 8 orders in the case at hand).

\subsubsection{$A_1$ and $A_2$}

Particles $A_1$ and $A_2$ also appear as $A_1A_1$ bound states.
However, there is no point in applying the complete bound state
quantization to them, because the error terms dominate over the higher
order contributions from \eqref{u_egyenlete}: the exponents of the
subleading finite size corrections $m_1$ and $\mu_{12}^2$ are smaller than $2\mu_{11}^1$ and $2\mu_{11}^2$. 
Nevertheless, the leading $\mu$-term can be verified by choosing suitable windows in $l$.

 In figures
\ref{fig:A1_energia} and \ref{fig:A2_energia} 
$\textrm{log}(|\Delta e|)$ is plotted against the prediction of
\eqref{elso_korrekcio} for $l=6..16$ and $l=6..22$. (the sign of
$\Delta e$ was found to be in accordance with 
\eqref{elso_korrekcio} for both $A_1$ and $A_2$)

In the case of $A_1$ perfect agreement is observed for $l=10..18$
in the sectors $I=0$ and $I=1$. For $I=2$ and $I=3$ the energy corrections
become too small and therefore inaccessible to TCSA (note that the
prediction for $I=3$ is of order $10^{-6}$).

In the case of $A_2$ precise agreement is found for $l=14..22$ in all
four sectors.

\subsubsection{$A_5$}

Here we present an interesting calculation that determines the leading mass
corrections of $A_5$. The standard formulas are inapplicable in
this case, because $m_5$ lies above the two-particle
threshold. However, it is instructive to consider the composition of
$A_5$ under the bootstrap principle and to evaluate the $\mu$-term prediction.

There are two relevant fusions
\begin{eqnarray*}
  A_1A_3\to A_5 \quad \textrm{ with }\quad \mu_{13}^5=0.2079 \\
  A_2A_2\to A_5 \quad \textrm{ with }\quad \mu_{22}^5=0.6581 
\end{eqnarray*}

\begin{figure}
% tomeg_5.plt
  \centering
\psfrag{igazi}{$m_5$}
\psfrag{l}{$l$}
\psfrag{e}{$e$}

\includegraphics[scale=1]{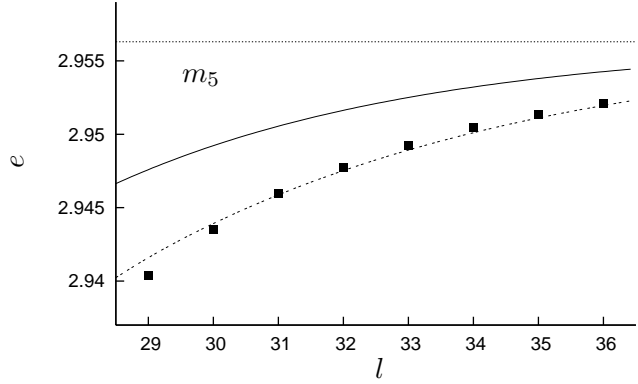}

\caption{
Finite size mass of $A_5$ as a function of the volume. The
squares represent the TCSA data which is compared to the leading
$\mu$-term (solid curve) and the solution of the quantization condition for the
``triple bound state`` $A_1A_1A_1$ (dotted curve). The straight line
shows the infinite volume mass.
\label{fig:A5}}
\end{figure}

Numerical evaluation of \eqref{mass_mu_term} shows  that the
contribution of the second fusion is negligible for 
$l>30$. The first fusion on the other hand yields 
a significant discrepancy when compared to TCSA data. 
 This failure 
is connected to the two-particle threshold and it can be explained in
terms of the bound state quantization.
Experience with $A_3$ suggests that one should
first take into account the energy corrections of $A_3$ and consider the
$A_1A_3\to A_5$ fusion afterwards. $A_3$ can be split into $A_1A_1$
leading  to the ''triple bound state'' $A_1A_1A_1\to A_5$. In 
infinite volume one has (see also fig. \ref{fusion} c.)
\begin{equation*}
  \ket{\theta}_5 \sim \ket{\theta-2i\bar{u}_{11}^3,\theta,\theta+2i\bar{u}_{11}^3}_{111}
\end{equation*}
The finite volume realization of this identification is most easily
carried out in the $I=0$ sector with
\begin{equation*}
  \ket{\{0\}}_{5,L} \sim \ket{\{0,0,0\}}_{111,L}
\end{equation*}
Setting up the three-particle Bethe-Yang
equations with rapidites $(iu,0,-iu)$:
\begin{eqnarray*}
  e^{-m_1\sin(u)L} S_{11}(iu)S_{11}(2iu)&=&1\\
   S_{11}(iu)S_{11}(-iu)&=&1\\
  e^{m_1\sin(u)L} S_{11}(-iu)S_{11}(-2iu)&=&1
\end{eqnarray*}
The second equation is automatically satisfied due to unitarity and
real analyticity, whereas
the first and the third are equivalent and they serve as a
quantization condition for $u$. 
The finite volume mass of $A_5$ is given in terms of the solution by
\begin{equation}
\label{A5_pred}
  m_5(l)=2\cos(u)+1
\end{equation}
In the large $L$ limit the infinite volume mass is reproduced by $u\to
2\bar{u}_{11}^3$. Figure  \ref{fig:A5} demonstrates the agreement
between TCSA and the prediction of \eqref{A5_pred}.

The possibility of solving the quantization of the triple bound state
in a moving frame looks very appealing. In  the general case the
rapidities are expected to take the form $(\theta_1+iu,\theta_2,\theta_1-iu)$ where
$\theta_1$ and $\theta_2$ do not necessarily coincide. However, the numerical precision of
our TCSA data was not sufficient to check our predictions.

\ssec{Multi-particle states}

\subsection{Bethe-Yang quantization in the bound state picture}

Multi-particle states in finite volume are denoted by 
\begin{equation*}
  \ket{\{I_1,\dots I_n\}}_{i_1i_2\dots i_n,L}
\end{equation*}
where the quantum numbers $I_j$ serve as an input to the Bethe-Yang
equations
\begin{equation}
  \label{Bethe}
Q_j(\theta_1,\dots,\theta_n)=m_{i_j}\sinh(\theta_j)L+\sum_{k\ne
  j}\delta _{i_ji_k}(\theta_j-\theta_k) =2\pi I_j\quad \quad j=1\dots n
\end{equation}
The energy  is calculated as
\begin{equation}
\label{multi_elsotipp}
  E=\sum_{j=1}^n m_{i_j}
  \cosh(\bar{\theta}_j)+\dots
\end{equation}
where $(\bar{\theta},\bar{\theta}_1,\dots,\bar{\theta}_n)$ is
the solution of \eqref{Bethe}. The dots indicate exponentially
decaying finite size
corrections. 

Let us consider a scattering state
$\ket{\{I,I_1,\dots,I_n\}}_{cb_1\dots b_n,L}$ composed of $n+1$
particles, the first one being  $A_c$.
 We determine the leading part of
the $\mu$-term by considering $A_c$ as an $A_aA_a$ bound state
\textit{inside the multi-particle state}. Therefore we write
\begin{equation}
\label{multi_azonositas}
   \ket{\{I,I_1,\dots,I_n\}}_{cb_1\dots b_n,L}
\sim
 \ket{\{I/2,I/2,I_1,\dots,I_n\}}_{aab_1\dots b_n,L}
\end{equation}
The energy is then determined by analytic continuation
of the $n+2$ particle Bethe-Yang equations. They read
\begin{eqnarray}
\label{multi_Bethe_1}
  e^{-m_a\sin(u)\cosh(\theta)L} e^{im_a \cos(u)\sinh(\theta)L}
  S_{aa}(2iu) \prod_{j=1}^n S_{ab_j}(\theta+iu-\theta_j) &=&1\\
\label{multi_Bethe_2}
  e^{m_a\sin(u)\cosh(\theta)L} e^{im_a \cos(u)\sinh(\theta)L}
  S_{aa}(-2iu) \prod_{j=1}^n S_{ab_j}(\theta-iu-\theta_j) &=&1\\
\label{multi_Bethe_3}
e^{im_{b_j}\sinh(\theta_j)L} S_{ab_j}(\theta_j-\theta-iu)
S_{ab_j}(\theta_j-\theta+iu) \mathop{\prod_{k=1}^n}_{k\ne j}
S_{b_jb_k}(\theta_j-\theta_k) &=& 1 
\end{eqnarray}
The ordinary $n+1$ particle Bethe-equations are reproduced in
the $L\to\infty$ limit by multiplying \eqref{multi_Bethe_1} and
\eqref{multi_Bethe_2} and making use of the bootstrap equation
\begin{equation*}
  S_{cb_j}(\theta)=S_{ab_j}(\theta+i\bar{u}_{ac}^a)S_{ab_j}(\theta-i\bar{u}_{ac}^a)
\end{equation*}

We now proceed similar to the previous section and
derive a formula for the leading correction. 
The shift in the imaginary part of the rapidity can be calculated by
making use of \eqref{multi_Bethe_1} and \eqref{S-matrix_polus} as
\begin{equation}
\label{multi-particle_u}
\Delta u=  u-\bar{u}_{ac}^a=\frac{\left(\Gamma_{aa}^c\right)^2}{2}
  e^{-\mu\cosh(\bar{\theta})L} e^{im_c\sinh(\bar{\theta})L/2}
  \prod_{j=1}^n S_{ab_j}(\bar{\theta}+i\bar{u}_{ac}^a-\bar{\theta}_j)
\end{equation}
Multiplying \eqref{multi_Bethe_1} and \eqref{multi_Bethe_2}
\begin{eqnarray}
\label{mB1}
 e^{i2m_a \cos(u)\sinh(\theta)L}
 \prod_{j=1}^n
 S_{ab_j}(\theta-iu-\theta_j)S_{ab_j}(\theta+iu-\theta_j) &=&1\\
\label{mB2}
e^{im_{b_j}\sinh(\theta_j)L} S_{ab_j}(\theta_j-\theta-iu)
S_{ab_j}(\theta_j-\theta+iu) \mathop{\prod_{k=1}^n}_{k\ne j}
S_{b_jb_k}(\theta_j-\theta_k) &=& 1 
\end{eqnarray}
Let us define
\begin{equation*}
  S_{ab_j}(\theta-iu-\theta_j)S_{ab_j}(\theta+iu-\theta_j)\approx 
S_{cb_j}(\theta-\theta_j)e^{i \Delta u \bar{\varphi}_{cb_j}(\theta-\theta_j)}
\end{equation*}
where
\begin{equation*}
  \bar{\varphi}_{cb_j}(\theta)=
i\varphi_{cb_j}(\theta+i\bar{u}_{ac}^a)-i\varphi_{cb_j}(\theta-i\bar{u}_{ac}^a)
\quad\quad\textrm{with}\quad\quad \varphi_{ab}(\theta)=\delta_{ab}'(\theta)
\end{equation*}
Using $2m_a \cos(u)\approx m_c-2\mu_{aa}^c \Delta u$ the logarythm
of \eqref{mB1} and \eqref{mB2} can be written as
\begin{eqnarray*}
Q_0(\theta,\theta_1,\dots,\theta_n)  &=&\left(2\mu_{aa}^c
   \sinh(\theta)L- \sum_{j=1}^n\bar{\varphi}_{cb_j}(\theta-\theta_j)\right) \Delta u \\
Q_j(\theta,\theta_1,\dots,\theta_n)  &=& 
\bar{\varphi}_{cb_j}(\theta-\theta_j)  \Delta u 
\end{eqnarray*}
The lhs. can be expanded around the $n+1$ particle solution
$(\bar{\theta},\bar{\theta}_n,\dots,\bar{\theta}_n)$ 
to arrive at
\begin{equation}
\label{rapieltolodas}
  \begin{pmatrix}
    \theta- \bar{\theta} \\ \theta_1- \bar{\theta}_1 \\ 
\vdots \\ \theta_n- \bar{\theta}_n
  \end{pmatrix}
=
\left(\mathcal{J}^{(n+1)}\right)^{-1} 
\begin{pmatrix}
  2\mu_{aa}^c
   \sinh(\bar{\theta})L-\sum_{j=1}^n
   \bar{\varphi}_{cb_j}(\bar{\theta}-\bar{\theta}_j)\\
\bar{\varphi}_{cb_1}(\bar{\theta}-\bar{\theta}_1) \\
\vdots \\
\bar{\varphi}_{cb_n}(\bar{\theta}-\bar{\theta}_n) 
\end{pmatrix}\Delta u 
\end{equation}
where
\begin{equation*}
  \mathcal{J}^{(n+1)}_{kl}=\frac{\partial Q_k}{\partial \theta_l}
\end{equation*}

The final result for the energy correction reads 
\begin{equation}
\label{multi_mu}
  \Delta E=-2\mu_{aa}^c  \cosh(\bar{\theta}) \Delta u +
  \begin{pmatrix}
    m_c\sinh(\bar{\theta}) \\ m_{b_1}\sinh(\bar{\theta}_1) \\
    \vdots \\ m_{b_n}\sinh(\bar{\theta}_n)
  \end{pmatrix}
\left(\mathcal{J}^{n+1}\right)^{-1} 
\begin{pmatrix}
  2\mu_{aa}^c
   \sinh(\bar{\theta})L-\sum_{j=1}^n
   \bar{\varphi}_{cb_j}(\bar{\theta}-\bar{\theta}_j)\\
\bar{\varphi}_{cb_1}(\bar{\theta}-\bar{\theta}_1) \\
\vdots \\
\bar{\varphi}_{cb_n}(\bar{\theta}-\bar{\theta}_n) 
\end{pmatrix}\Delta u 
\end{equation}
with $\Delta u$ given by \eqref{multi-particle_u}.

Based on the previous section it is expected that there is a similar
contribution for every fusion leading to each one of the constituents of
the multi-particle state. 

\subsection{Multi-particle states -- Numerical analysis}

We first consider finite size corrections to $A_1A_3$ states. They are
not the lowest lying two-particle states in the spectrum, but they possess
the largest $\mu$-term which is connected to the $A_1A_1\to A_3$
fusion. Given a particular state $\ket{\{I_1,I_3\}}_{13,L}$  the following procedure
is performed at each value of the volume:
\begin{itemize}
\item The two-particle Bethe-Yang equation for
  $\ket{\{I_1,I_3\}}_{13,L}$ is solved and the
  $\mu$-term is calculated according to \eqref{multi_mu}.
\item The exact three-particle Bethe-Yang equation is solved for $\ket{\{I_1,I_3/2,I_3/2\}}_{111,L}$
\end{itemize}
The results for different $A_1A_3$ levels are shown in figure
\ref{fig:A1A3_enkorr}. The situation is similar to 
 the case of the $A_3$ one-particle levels: 
the  bound state quantization yields a remarkably accurate
prediction, whereas
the single $\mu$-term
prediction only becomes correct in the $L \to\infty$ limit.

In table \ref{tab:en_A1A3_1} we present a numerical example for the
dissociation of the bound state inside the two-particle
state. In this case an $A_1A_3$ state turns into a conventional $A_1A_1A_1$
three-particle state at $l_c\approx 30$. 

Finite size corrections to $A_1A_1$ and $A_1A_2$
states are also investigated, the leading $\mu$-term given by the fusions $A_1A_1\to A_1$ and
$A_1A_1\to A_2$, respectively. In the former case we 
calculate separately the contribution associated to both $A_1$ particles and
 add them to get the total correction.
Results are exhibited in figures
\ref{fig:A1A1_enkorr} and \ref{fig:A1A2_enkorr} and formula
\eqref{multi_mu} is verified in both cases. 

\ssec{Finite volume form factors}

The connection between finite volume and infinite
volume form factors was derived in  \cite{fftcsa1} as
\begin{eqnarray}
\nonumber
_{j_1\dots j_m,L}  \bra{\{I'_1,\dots,I'_m\}}
\mathcal{O}(0,0)\ket{\{I_1,\dots,I_n\}}_{i_1\dots i_n,L}=\\
\label{fvff}
\frac{F^{\mathcal{O}}(\bar{\theta}'_m+i\pi,\dots,\bar{\theta}'_1+i\pi,\bar{\theta}_1,\dots,\bar{\theta}_n)_{j_m\dots j_1i_1\dots i_n}}
{\sqrt{
\rho_{i_1\dots i_n}(\bar{\theta}_1,\dots,\bar{\theta}_n)
\rho_{j_1\dots j_m}(\bar{\theta}'_1,\dots,\bar{\theta}'_m)}}
+ O(e^{-\mu' L})
\end{eqnarray}
where the rapidities $\bar{\theta}$ are solutions of the corresponding Bethe-Yang
equations and
it is supposed that $\bar{\theta}_j\ne \bar{\theta}'_k$ whenever
$i_j=i_k$. The extension of \eqref{fvff} to include disconnected
terms can be found in \cite{fftcsa2}. The proportionality factor in \eqref{fvff} is
given by the density of states 
\begin{equation*}
  \rho_{i_1\dots i_n}^{(n)}(\bar{\theta}_1,\dots,\bar{\theta}_n)=\textrm{det}
  \mathcal{J}^{(n)} \quad ,\quad 
 \mathcal{J}^{(n)}_{kl}=\frac{\partial Q_k}{\partial
   \theta_l}\quad,\quad
k,l=1\dots n
\end{equation*}
which can be used to identify formally the finite volume and infinite
volume states as
\begin{equation*}
   \ket{\{I_1,\dots I_n\}}_{i_1i_2\dotsi_n,L}\sim
\frac{1}{\sqrt{ \rho_{i_1\dots i_n}^{(n)}(\bar{\theta}_1,\dots,\bar{\theta}_n)}}
 \ket{\bar{\theta}_1,\dots,\bar{\theta}_n}_{i_1\dots i_n}
\end{equation*}

Based on general arguments
it was shown in  \cite{fftcsa1} that $\mu'\ge \mu$ where $\mu$ is
determined by the pole of the S-matrix closest to the physical
line. A systematic finite volume perturbation theory (Lüscher's method
applied to form factors)
is not available. However, it is expected that
the actual value of $\mu'$ depends on what diagrams contribute
to the form factor in question. Apart from the insertion of the local
operator they coincide with the diagrams determining the finite size
corrections of the multi-particle state. Therefore $\mu'$ is
associated to the bound state structure of the constituents 
of the multi-particle state. 
In this section it is shown that the leading correction term can be
obtained by the bound state quantization.

\subsection{Elementary one-particle form factors}

\eqref{fvff} yields a simple prediction for the elementary one-particle form factor:
\begin{equation}
\label{egyreszecskes_elsotipp}
  F^{\mathcal{O}}_c(I,L)\equiv\bra{0}\mathcal{O}(0,0)\ket{\{I\}}_{c,L}=\frac{F_c^\mathcal{O}}{\sqrt{EL}}+
O(e^{-\mu' L})
\end{equation}
where $E$
is the one-particle energy, and
$F_c^\mathcal{O}=F_c^\mathcal{O}(\theta)$ is the infinite volume
one-particle form factor, which is constant by 
Lorentz symmetry. 

The $\mu$-term associated to \eqref{egyreszecskes_elsotipp} is derived
by employing the bound state quantization. We gain some intuition from the previous sections where it was
found that the bound state $A_aA_a$ may dissociate at a critical volume $L_c$.
For $L<L_c$ there is no one-particle level of type $A_c$ in the
given sector of the spectrum, however an $A_aA_a$ scattering state
appears instead. Finite volume form factors of this state are calculated using \eqref{fvff} as
\begin{equation}
\label{ketreszecskes_fazisnelkul}
    F^{\mathcal{O}}_c(I,L)=
\frac{F^{\mathcal{O}}({\theta}_1,{\theta}_2)_{aa}}
{\sqrt{\rho_{aa}({\theta}_1,{\theta}_2)}}
\quad \textrm{for } L<L_c
\end{equation}

The generalization to $L>L_c$ seems to be straightforward: one has to
 continue analytically \eqref{ketreszecskes_fazisnelkul} to the
solutions of the Bethe-Yang equation with imaginary rapidities
$\theta_{1,2}=\theta\pm iu$. However, note that equations
\eqref{egyreszecskes_elsotipp} and \eqref{ketreszecskes_fazisnelkul}
are valid up to a phase factor. 
In order to
continue analytically to imaginary rapidities we also need to fix
this phase\footnote{The phase of a (nondiagonal) infinite volume form factor is
  unphysical in the sense that it may be redefined by a complex
  rotation of the state vectors and 
physical quantities, e.g. correlation
functions, do not depend on such redefinitions.
 However, the bootstrap program uniquely assigns a phase
  to each form factor. }.

The two-particle form factor satisfies
\begin{equation*}
  F^{\mathcal{O}}(\theta_1,\theta_2)_{aa}=S_{aa}(\theta_1-\theta_2)F^{\mathcal{O}}(\theta_2,\theta_1)_{aa}
\end{equation*}
The simplest choice for the phase is therefore
\begin{equation}
\label{fazisvalasztas}
  F^{\mathcal{O}}(\theta_1,\theta_2)_{aa}=\sqrt{S_{aa}(\theta_1-\theta_2)}
\left|F^{\mathcal{O}}(\theta_1,\theta_2)_{aa}\right|
\end{equation}
This choice is dictated by CPT symmetry
\cite{maiani_testa_final_state_interaction,lellouch_luscher,k_to_pipi},
and it is respected by all known solutions of the 
form factor bootstrap axioms. 
There is a sign ambiguity caused by the square root, but it can
be fixed by demanding $(S_{aa}(0))^{1/2}=i$ and continuity.
Using \eqref{fazisvalasztas} 
\begin{equation*}
    F^{\mathcal{O}}_c(I,L)=
\frac{\sqrt{S_{aa}(\theta_2-\theta_1)} 
F^{\mathcal{O}}(\theta_1,\theta_2)_{aa}}
{\sqrt{\rho_{aa}(\theta_1,\theta_2)}}
\end{equation*}
and upon analytic continuation 
\begin{equation}
  \label{analitikus_elfolytatas}
    F^{\mathcal{O}}_c(I,L)=
\frac{\sqrt{S_{aa}(-2iu)} 
F^{\mathcal{O}}(\theta+iu,\theta-iu)_{aa}}
{\sqrt{\rho_{aa}(\theta+iu,\theta-iu)}}
\quad \textrm{for } L>L_c
\end{equation}

It is easy to see that the result
\eqref{egyreszecskes_elsotipp} is reproduced in the $L\to\infty$ limit.
First observe that 
\begin{equation*}
  F^{\mathcal{O}}(\theta+iu,\theta-iu)_{aa}\sim 
\frac{\Gamma_{aa}^c}{2(u-\bar{u}^c_{aa})}F_c^{\mathcal{O}}(\theta)
\end{equation*}
The residue of $\rho_{aa}$ is determined by $\varphi_{aa}(2iu)$
and it reads
\begin{equation*}
   \rho(\theta+iu,\theta-iu)_{aa}\sim
2m_aL \cos(u)\cosh(\theta) (-i) \frac{S_{aa}'(2iu)}{S_{aa}(2iu)}=
m_cL \cosh(\theta) \frac{1}{2(u-\bar{u}_{ac}^a)}
\end{equation*}
The singularities in the numerator and denominator of
\eqref{analitikus_elfolytatas} cancel  and  indeed
\begin{equation*}
  F^{\mathcal{O}}_c(I,L)\sim
\frac{F_c^{\mathcal{O}}}{ \sqrt{m_cL\cosh(\theta)}}
\end{equation*}
We emphasize that it is
crucial to include the extra normalisation factor
$\sqrt{S_{aa}(-2iu)}$ to obtain a meaningful result.

Expression \eqref{analitikus_elfolytatas} can be developed into a
Taylor-series in $u-\bar{u}_{ac}^a$. The first order correction is
evaluated in Appendix A and it reads
\begin{eqnarray}
 \nonumber
 && F^{\mathcal{O}}_c(I,L)=
\frac{F_c^{\mathcal{O}}}{\sqrt{E_c^0L}}
-\frac{2i\Gamma_{aa}^c \left(F^{\mathcal{O}}_{aa}\right)'}{ \sqrt{E_c^0L}} (u-\bar{u}_{ac}^a)+
\\
&&
+\frac{F_c^{\mathcal{O}}}{ \sqrt{E_c^0L}}
\left[
\frac{2S_{aa}^{c,0} }{\left(\Gamma_{aa}^c \right)^2}
+\frac{m_c\mu}{(E_c^0)^2}-
\left(\frac{m_a^2}{m_c^2}E_c^0-\frac{\mu^2}{E_c^0}\right)L
 \right](u-\bar{u}_{ac}^a) +O(e^{-2\mu L})
\label{ff_mu_tag_2}
\end{eqnarray}
where
\begin{eqnarray*}
  \left(F^{\mathcal{O}}_{aa}\right)'&=&\lim_{\theta-\theta'=-2i\bar{u}_{ac}^a}
\frac{d}{d\theta}   F^{\mathcal{O}}(\theta,\theta')_{aa}\\
S_{aa}^{c,0}&=&\lim_{u\to\bar{u}_{ac}^a}\left(
 S_{aa}(2iu)-\frac{\left(\Gamma_{aa}^c \right)^2}{2(u-\bar{u}_{ac}^a)}\right)
\end{eqnarray*}
$E_c^0$ is the ordinary one-particle energy and the rapidity shift
$u-\bar{u}_{ac}^a$ is given by \eqref{uminusubar}. 

\subsection{One-particle form factors -- Numerical analysis}

Let us introduce the dimensionless form factors as
\begin{equation*}
  f_i(I,l)=\frac{\bra{0}\eps(0,0)\ket{\{I\}}_{i,L}}{m_1}
\end{equation*}
The machinery of \cite{fftcsa1} is used to determine
$f_i(I,l)$ for $i=1,2,3$ and $I=0,1,2,3$. The numerical results are
compared to the exact infinite volume form factors \cite{ising_ff1,ising_ff2}. 

\begin{figure}
% f1_3.plt
  \centering
\small
\psfrag{Tspin0}{$\bra{0}\eps\ket{\{0\}}_3$}
\psfrag{Tspin1}{$\bra{0}\eps\ket{\{1\}}_3$}
\psfrag{Tspin2}{$\bra{0}\eps\ket{\{2\}}_3$}
\psfrag{Tspin3}{$\bra{0}\eps\ket{\{3\}}_3$}
\psfrag{l}{$l$}
\psfrag{exact}{$F_3$}
\psfrag{f3}{$|\bar{f}_3|$}
  \includegraphics[scale=0.5,angle=-90]{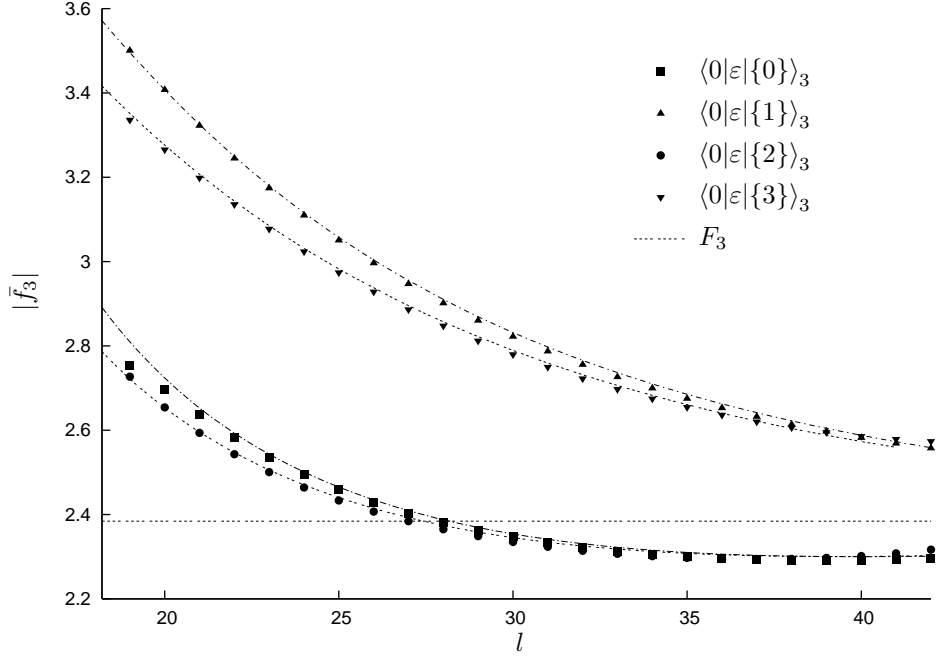}
\caption{
Elementary finite volume form factors of $A_3$ one-particle
levels. 
Here the normalization  \eqref{F3normalas} is applied to
obtain a finite $l\to\infty$ limit, which is given by the infinite
volume form factor $F_3=\bra{0}\eps\ket{A_3(\theta)}$. The TCSA 
data are plotted against the bound state prediction. The ordinary
evaluation of $\bar{f}_3$ is simply the constant $F_3$.
\label{fig:F3}}
\end{figure}

We start our investigation with  $f_3(I,l)$, for which
relatively large exponential corrections were already
reported in \cite{fftcsa1}. 
It is convenient to consider
\begin{equation}
\label{F3normalas}
  \bar{f}_3(I,l)=\left(e_0l\right)^{1/2} f_3(I,L)\quad\textrm{with}
\quad \lim_{l\to\infty}\bar{f}_3(I,l)=F_3
\end{equation}
The numerical results are demonstrated in
fig. \ref{fig:F3}. Note, that this is exactly the same figure as
4.6.~(c) in \cite{fftcsa1}, but this time the interpretation
of the huge deviations from $F_3$ is also provided. 

We also tried to verify the predictions  for $f_1$
and $f_2$. In the latter case reasonably good agreement was found
with TCSA, the results are demonstrated in fig. \ref{fig:F2}. In the case
of $f_1$ we encountered the unpleasant situation that the F-term
decays slower than the TCSA errors grow, thus making the observation of the
$\mu$-term impossible.

\begin{figure}
% f2_1_3.plt, f2_2_3.plt, f2_3_3.plt
  \centering
\tiny
\psfrag{l}{$l$}
\psfrag{f13}{$f_{13}$}
\psfrag{f23}{$f_{23}$}
\psfrag{f33}{$f_{33}$}
\subfigure[$_{1}\bra{\{0\}}\eps\ket{\{3\}}_{3,L}$]{\includegraphics[scale=0.4,angle=-90]{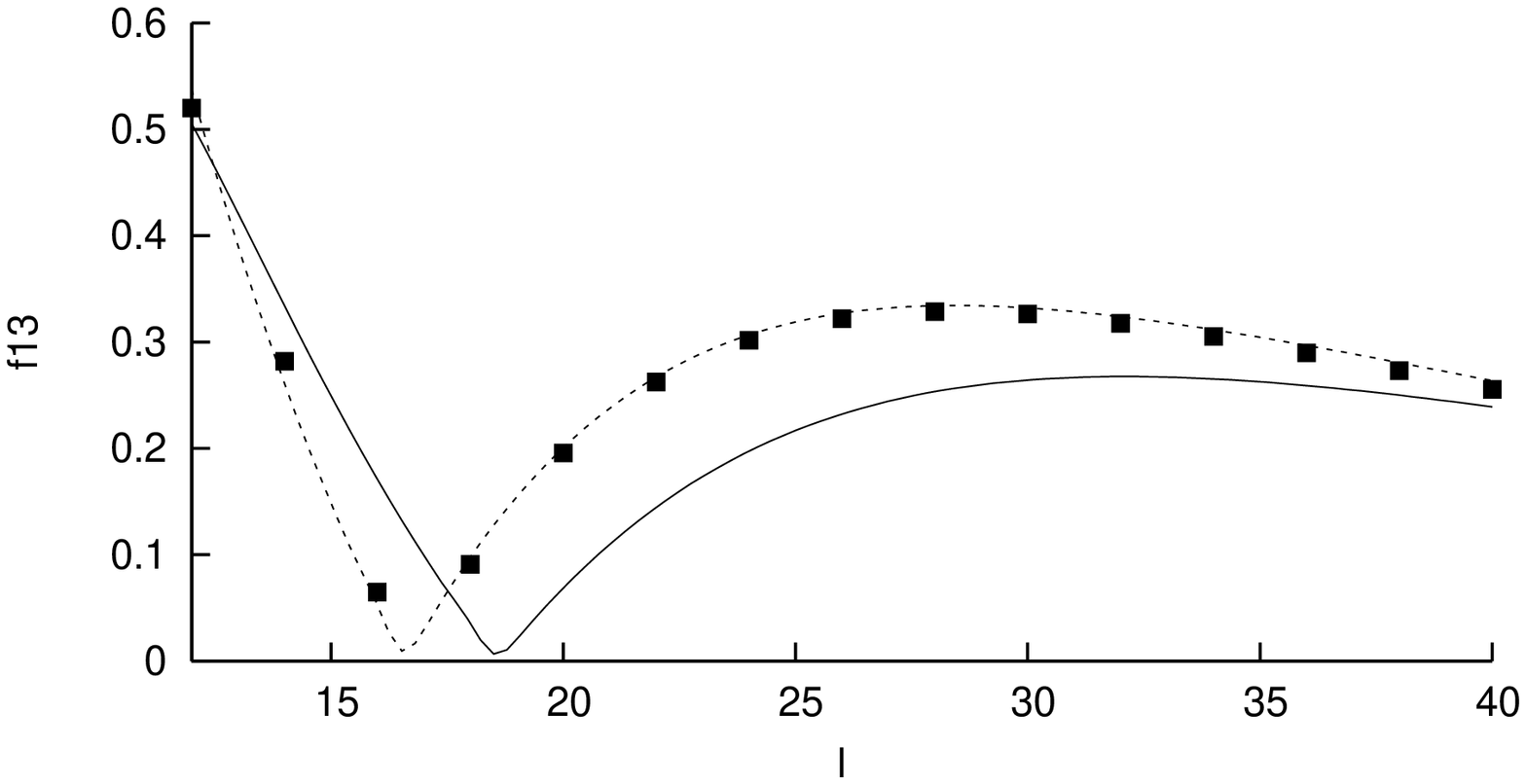}}
\subfigure[$_{1}\bra{\{0\}}\eps\ket{\{1\}}_{3,L}$]{ \includegraphics[scale=0.4,angle=-90]{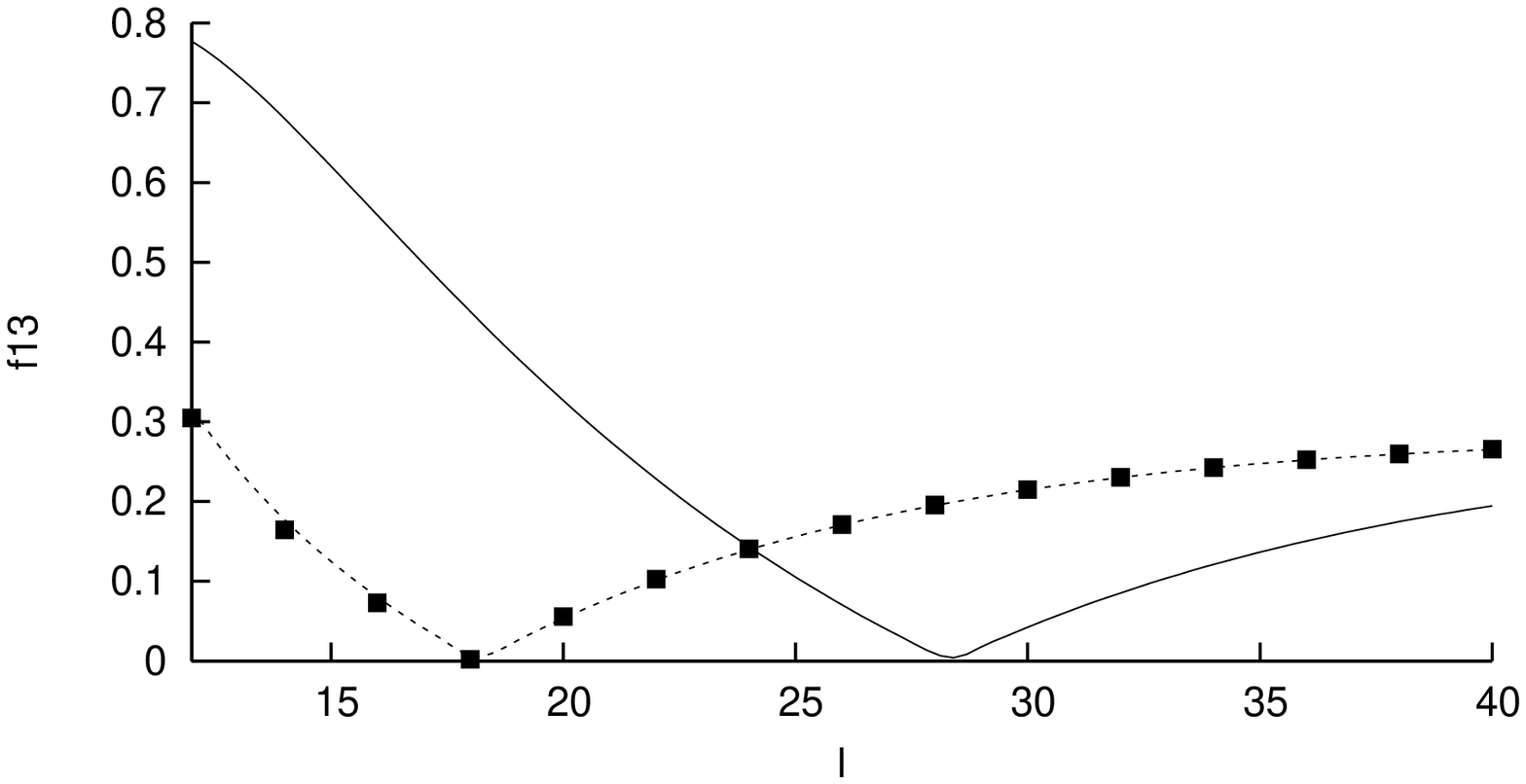}}

\subfigure[$_{2}\bra{\{-2\}}\eps\ket{\{3\}}_{3,L}$]{ \includegraphics[scale=0.4,angle=-90]{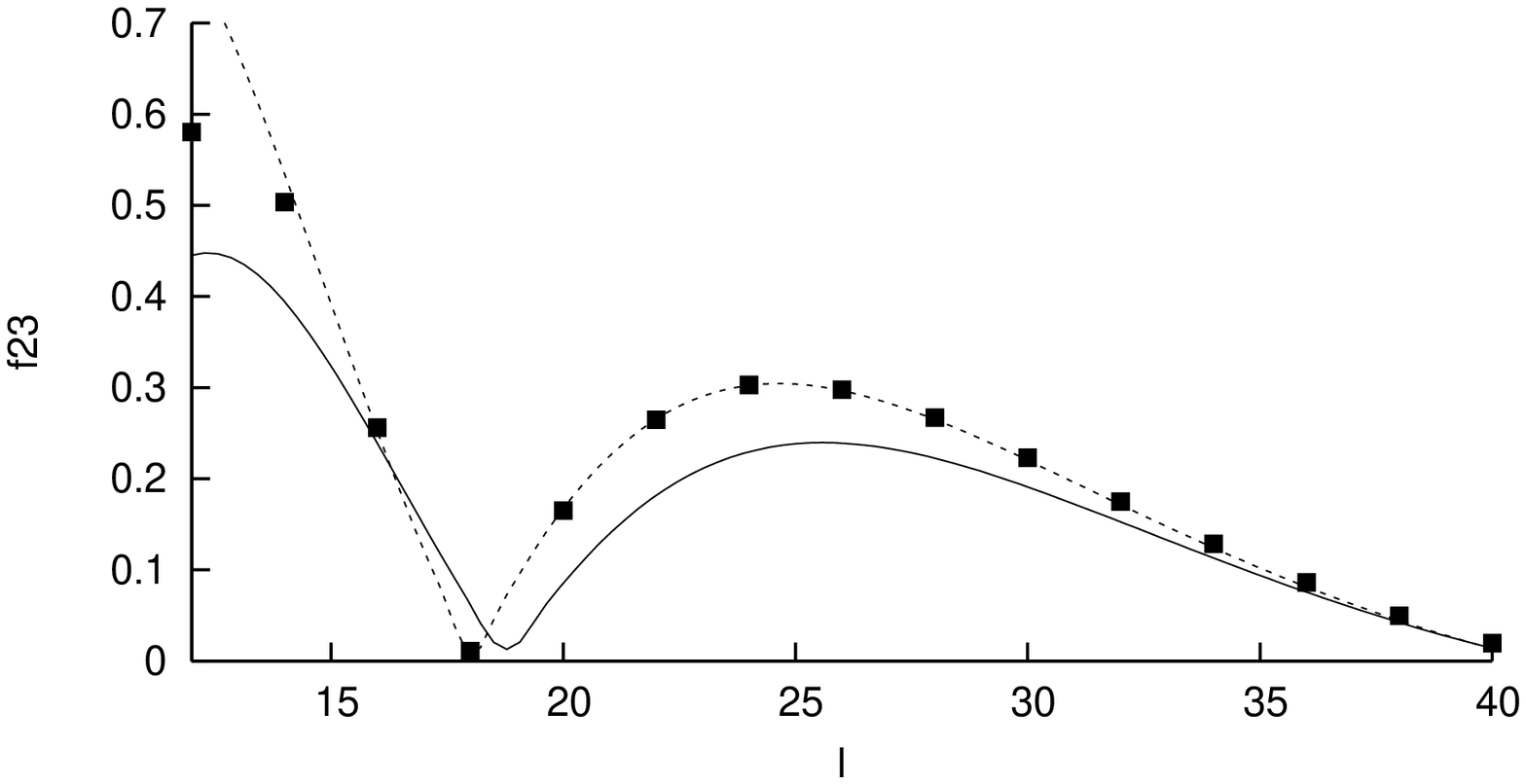}}
\subfigure[$_{3}\bra{\{-3\}}\eps\ket{\{3\}}_{3,L}$]{\includegraphics[scale=0.4,angle=-90]{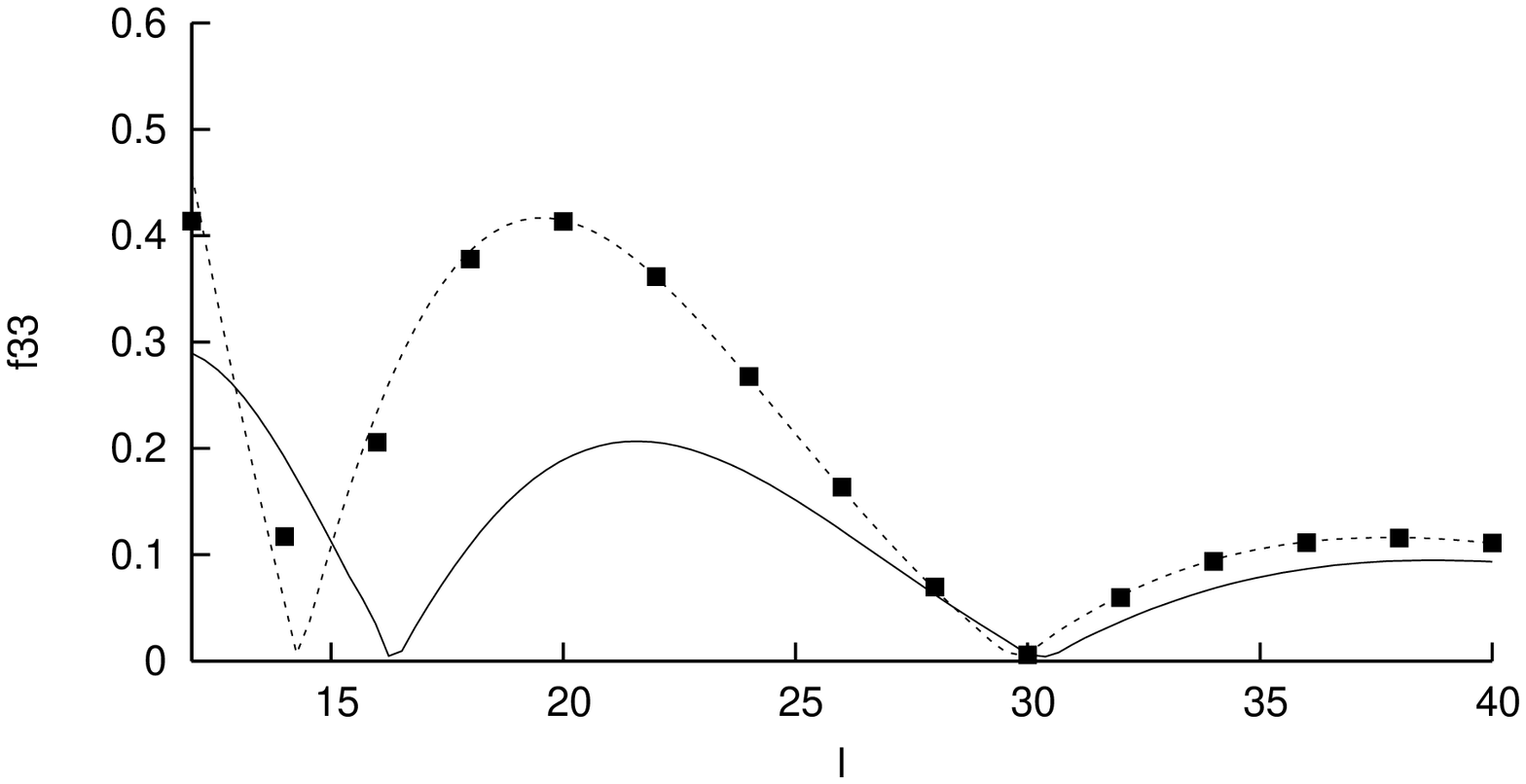}}
\caption{One-particle--one-particle form factors, dots correspond to
  TCSA data. The solid lines represent the ordinary evaluation of the finite volume form
  factors, while the dotted lines show the bound state
  prediction.  
\label{fig:Fx3}}
\end{figure}

It is straightforward to generalize \eqref{analitikus_elfolytatas} to
matrix elements between two different one-particle
 states. For $b\ne c$ one has for example
\begin{equation*}
_b\bra{\{I_b\}}\eps\ket{\{I_c\}}_{c,L}=\frac{F^{\eps}(\theta_b+i\pi,\theta+iu,\theta-iu)_{baa}}
{\sqrt{\rho_b(\theta_b)\rho_{aa}(\theta+iu,\theta-iu)}}
\end{equation*}
Numerical examples are presented in figures \ref{fig:Fx3} (a)-(c) for $c=3$ and
$b=1,2$. 

The most interesting case is the one shown in
fig. \ref{fig:Fx3} (d) where the matrix element between
two different $A_3$ one-particle states are investigated. This can be done by considering
both $A_3$ particles as the appropriate $A_1A_1$ bound states and then
calculating the finite volume form factor
$_3\bra{\{I\}}\eps\ket{\{I'\}}_{3,L}$ as
\begin{eqnarray*}
 _{11}\bra{\{I/2,I/2\}}\eps\ket{\{I'/2,I'/2\}}_{11,L}=
\frac{F^\eps(\theta+iu+i\pi,\theta-iu+i\pi,\theta'+iu',\theta'-iu')_{1111}}
{\sqrt{\rho_{11}(\theta'+iu',\theta'-iu')\rho_{11}(\theta+iu,\theta-iu)}}
\end{eqnarray*}
Once again we find complete agreement with the TCSA data.

\subsection{Elementary multi-particle form factors}

The generalization of \eqref{analitikus_elfolytatas} to multi-particle
states is straightforward, the only task is to find the 
appropriate phase factor.
Similar to the one-particle case one has
\begin{equation*}
  F^{\mathcal{O}}(\theta_1,\dots,\theta_m)_{b_1\dots b_m}=
\sqrt{\prod_{i<j} S_{b_ib_j}(\theta_i-\theta_j)}
\left|F^{\mathcal{O}}(\theta_1,\dots,\theta_m)_{b_1\dots b_m}\right|
\end{equation*}
A general $n$ particle finite volume form factor with real
rapidities can thus be
written as
\begin{eqnarray*}
  \sqrt{\frac{\prod_{i<j} S_{b_ib_j}(\theta_j-\theta_i)}{\rho^{n}
({\theta}_1,\dots,{\theta}_{n})_{b_1\dots
  b_{n}}}}
 F^{\mathcal{O}}(\theta_1,\dots,\theta_{n})_{b_1\dots b_{n}}
\end{eqnarray*}
Substituting the solution of the Bethe-equation for the state
$\ket{\{I/2,I/2,I_1,\dots,I_n\}}_{aab_1\dots b_n,L}$ and  
making use of the real analycity condition
\begin{equation*}
 |S_{ab_j}(\theta_j-\theta-iu)S_{ab_j}(\theta_j-\theta+iu)|=1 
\end{equation*}
one gets
\begin{equation}
\label{multi_particle_form_factor} 
\bra{0}\mathcal{O}\ket{\{I/2,I/2,I_1,\dots,I_n\}}_{aab_1\dots b_n,L}=
\frac{\sqrt{S_{aa}(-2iu)}
\left| F^{\mathcal{O}}(\theta+iu,\theta-iu,\theta_1,\dots,\theta_{n})_{aab_1\dots b_{n}}
\right|
}{\sqrt{\rho^{(n+2)}
(\theta+iu,\theta-iu,\theta_1,\dots,\theta_{n})_{aab_1\dots
  b_{n}}}}
\end{equation}
up to a physically irrelevant phase.

It is easy to show once again that the ''naive'' result is reproduced in the
$L\to\infty$ limit. To do so, we first quote the dynamical pole equation of the infinite volume form factor:
\begin{equation*}
  F^{\mathcal{O}}({\theta}+iu,{\theta}-iu,{\theta}_1,\dots,{\theta}_{n})
_{aab_1\dots b_{n}}=\frac{\Gamma_{aa}^c}{2(u-\bar{u}_{ac}^a)}
F^{\mathcal{O}}({\theta},{\theta}_1,\dots,{\theta}_{n})
_{cb_1\dots b_{n}}+O(1)
\end{equation*}
The singularity of $\rho^{(n+2)}$ is given by
\begin{equation*}
\textrm{Res}_{u\to \bar{u}_{ac}^a} \rho^{(n+2)}
(\theta+iu,\theta-iu,\theta_1,\dots,\theta_n)_{aab_1\dots b_n}=\frac{1}{2}\rho^{(n+1)}
(\theta,\theta_1,\dots,\theta_n)_{cb_1\dots b_n}  
\end{equation*}
The ''naive'' formula is now recovered by inserting the last two
equations into \eqref{multi_particle_form_factor}. 

The leading exponential corrections can be obtained by plugging
\eqref{multi-particle_u} and \eqref{rapieltolodas} into
\eqref{multi_particle_form_factor} and expanding to first order in  $u-\bar{u}_{ac}^a$.
This procedure is straightforward
but quite lengthy, therefore we refrain from giving the details of the calculations.

\begin{figure}
% f3_11_1.plt
  \centering
\psfrag{l}{$l$}
\psfrag{f111}{$f_{13}$}
\subfigure[$\bra{0}\eps\ket{\{1,1\}}_{13,L}$]{ \includegraphics[scale=0.5,angle=-90]{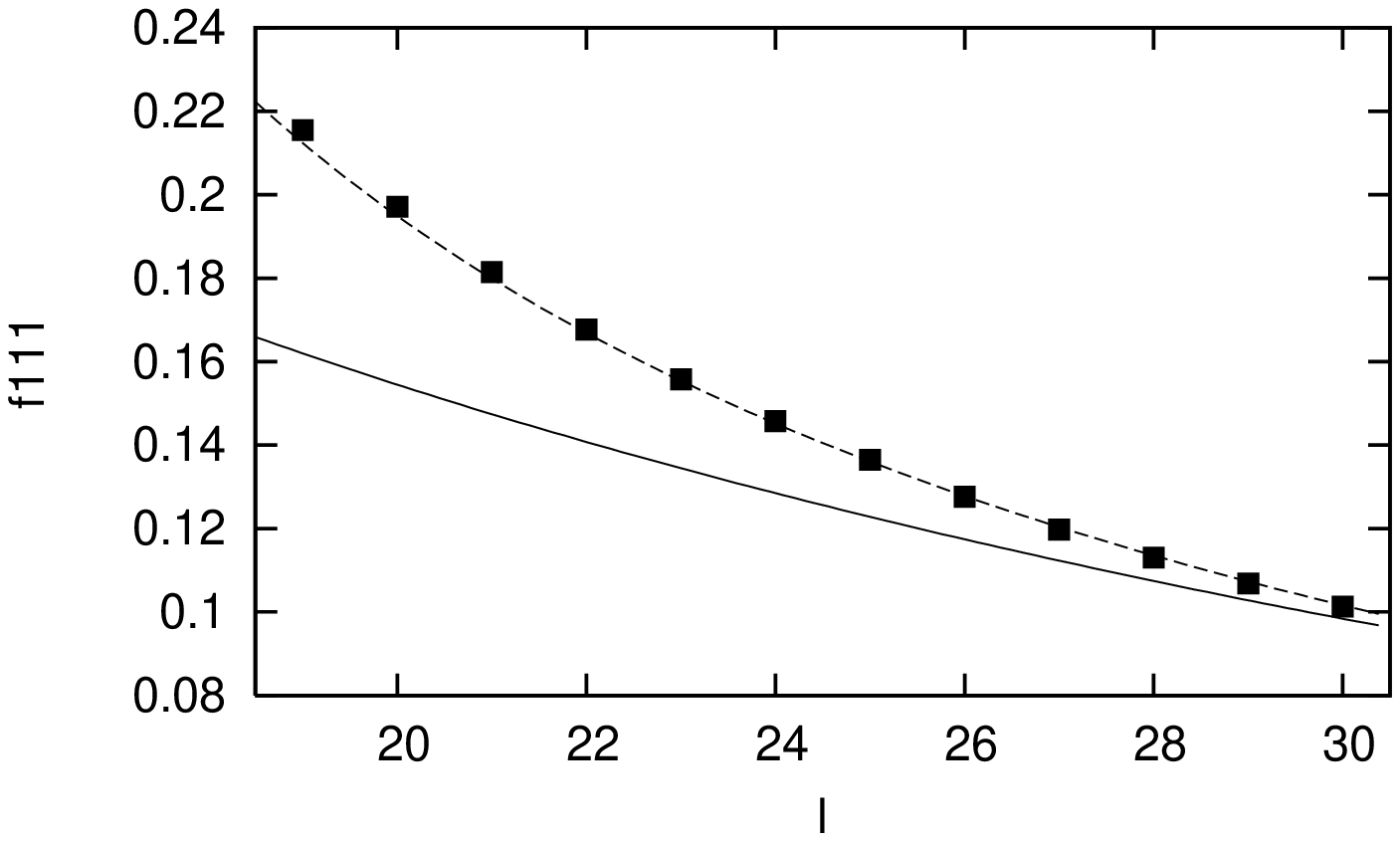}}
\subfigure[$\bra{0}\eps\ket{\{2,0\}}_{13,L}$]{ \includegraphics[scale=0.5,angle=-90]{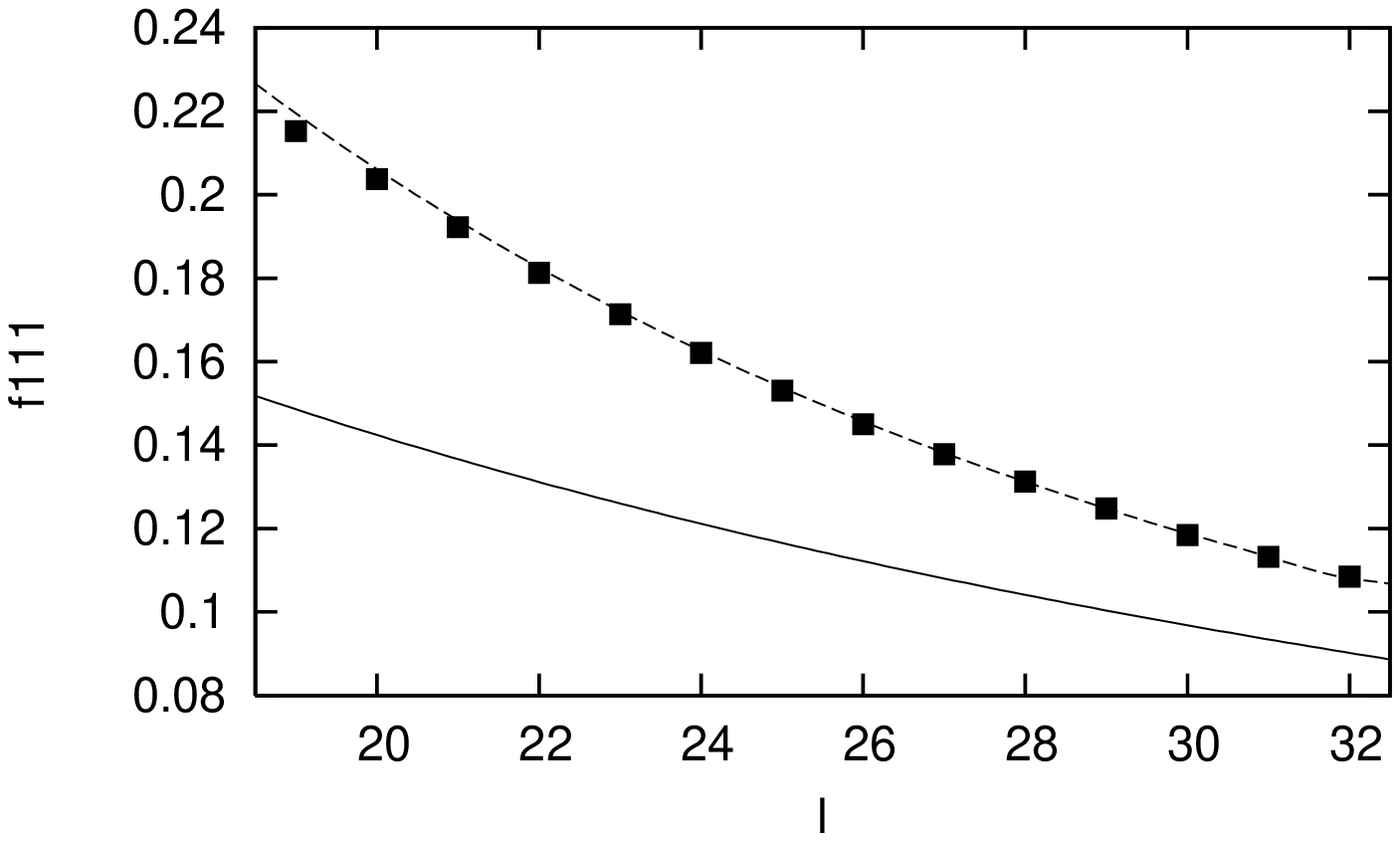}}
\caption{
Elementary form factors of $A_1A_3$ scattering states, dots correspond
to TCSA data. The solid lines are obtained by a ``naive``
evaluation of the finite volume form factors, while the dotted line
represents the bound state prediction. (in this case $A_1A_1A_1$ form
factors at the appropriate rapidities)
\label{fig:F13_durch_F111}}
\end{figure}

In fig. \ref{fig:F13_durch_F111} two examples are presented for the
evaluation of \eqref{multi_particle_form_factor} applied to $A_1A_3$
two-particle states. 

\ssec{Conclusions}

In this work we determined the generalization of L\"uscher's
$\mu$-term associated to moving
one-particle states, arbitrary scattering states and finite volume
form factors. Our method is based on the bootstrap principle of the infinite
volume theory  which
states that a particle which is a bound state of two others is indeed
indistinguishable from the two-particle state with the appropriate
(imaginary) momenta. An analytic continuation of the
Bethe-Yang equations was used to quantize the bound states in finite volume. 

The analytic results were tested by comparing the predictions to
high-precision TCSA data of the Ising model; a satisfactory agreement
was observed in each case.  We also demonstrated that the bound state
quantization goes beyond the leading mu-term corresponding to the
fusion and gives a resummation of all powers of $e^{-\mu_{aa}^cL}$. This
is a substantial improvement over the inclusion of the leading Lüscher
term when other sources of finite volume corrections (other fusions,
F-terms) can be neglected.  Our main example was provided by particle
$A_3$ from the E$_8$ scattering theory.

As an analytic check  the calculations were compared to
 the leading order results of the exact TBA equations.
The agreement between the two approaches was explicitly demonstrated
in the case of the one-particle levels of the Lee-Yang model. 

Our results can also be applied in nonintegrable models for states
below the first inelastic threshold. 
However, the calculations only apply to symmetric fusions of the type
$A_aA_a\to A_c$. The nonsymmetric case requires an extension  of the
Bethe-Yang equations to incorporate multi-channel scattering. 
Such a quantization scheme could also be used to describe
finite volume states in nonintegrable theories above the two-particle
threshold.

\subsection*{Acknowledgements}

The author is grateful to G. Tak\'acs for his support during the
completion of this work and for correcting the manuscript. The author
would also like to thank
Z. Bajnok for interesting discussions. This research was
partially supported by the Hungarian research fund OTKA K60040.

\appendix

\renewcommand{\theequation}{\Alph{section}.\arabic{equation}}
\renewcommand{\thetable}{\Alph{section}.\arabic{table}}
\renewcommand{\thefigure}{\Alph{section}.\arabic{figure}}

\ssec{$\mu$-term for the one-particle form factor}

Here we develop the first order correction to \eqref{analitikus_elfolytatas}.
Using  the exchange axiom 
\begin{equation}
 \label{analitikus_elfolytatas_2}
    F^{\mathcal{O}}_c(I,L)=
\frac{
F^{\mathcal{O}}(\theta-iu,\theta+iu)_{aa}}
{\sqrt{S_{aa}(-2iu)\rho_{aa}(\theta+iu,\theta-iu)}}
\end{equation}
The form factor axioms imply that
\begin{equation*}
  \lim_{u\to \bar{u}_{ac}^a}F^{\mathcal{O}}(\theta-iu,\theta+iu)_{aa}  =
\frac{1}{\Gamma_{aa}^c}F_c^{\mathcal{O}}
\end{equation*}
The simple pole of $\varphi_{aa}(2iu)$ in $\rho_{aa}$ is cancelled by
$S_{aa}(-2iu)$, therefore both the numerator and the denominator of
\eqref{analitikus_elfolytatas_2} have continuous limits as $u\to
\bar{u}_{ac}^a$. 

The form factor $F^{\mathcal{O}}(\theta-iu,\theta+iu)_{aa}$ only
depends on $u$ by Lorentz-symmetry. Therefore
\begin{equation*}
  F^{\mathcal{O}}(\theta-iu,\theta+iu)_{aa}=
\frac{1}{\Gamma_{aa}^c}
F_c^{\mathcal{O}}-2i\left(F^{\mathcal{O}}_{aa}\right)'(u-\bar{u}_{ac}^a)+
\dots
\end{equation*}
where
\begin{equation*}
  \left(F^{\mathcal{O}}_{aa}\right)'=
\frac{d}{d\theta}   F^{\mathcal{O}}(\theta,\theta')_{aa}
\Big|_{\theta-\theta'=-2i\bar{u}_{ac}^a}
\end{equation*}
Expanding the S-matrix element into a Laurent-series in the vicinity
of the pole 
\begin{eqnarray*}
S_{aa}(2iu)&=&\frac{\left(\Gamma_{aa}^c \right)^2}{2(u-\bar{u}_{ac}^a)}+
S_{aa}^{c,0}+\dots\\
S_{aa}(-2iu)&=&\frac{2(u-\bar{u}_{ac}^a)}{\left(\Gamma_{aa}^c
  \right)^2}-
\left(\frac{2(u-\bar{u}_{ac}^a)}{\left(\Gamma_{aa}^c \right)^2}\right)^2
S_{aa}^{c,0}+\dots
\end{eqnarray*}
Expanding the denominator:
\begin{eqnarray*}
&&  S(-2iu)\rho_{aa}(\theta+iu,\theta-iu)=\\
&&S(-2iu)E_1E_2L^2-i(E_1+E_2)L S'(2iu)\Big(S(-2iu)\Big)^2=\\
&&\frac{E_c L}{\left(\Gamma_{aa}^c \right)^2} +
\left(\frac{2E_1E_2L^2}{\left(\Gamma_{aa}^c \right)^2}
-\frac{4E_cL}{\left(\Gamma_{aa}^c \right)^4}S_{aa}^{c,0}\right) 
(u-\bar{u}_{ac}^a)+\dots
\end{eqnarray*}
where 
\begin{eqnarray*}
E_c=E_1+E_2=2m_a\cos(u)\cosh(\theta)
\end{eqnarray*}
Putting all this together
\begin{eqnarray}
 \nonumber
 && F^{\mathcal{O}}_c(I,L)=
\frac{F_c^{\mathcal{O}}}{ \sqrt{E_cL}}+\\
\nonumber
 &&+\left[\frac{-2i\Gamma_{aa}^c \left(F^{\mathcal{O}}_{aa}\right)'}{ \sqrt{E_cL}}
+
\frac{F_c^{\mathcal{O}}}{ \sqrt{E_cL}^3}\left(-E_1E_2L^2+
\frac{2E_cL}{\left(\Gamma_{aa}^c \right)^2}S_{aa}^{c,0} 
\right) \right] (u-\bar{u}_{ac}^a)+\dots
\end{eqnarray}
Note that in the preceding formulas $E_c$ does include the leading order correction to
the usual one-particle energy $E_c^0=\sqrt{m_c^2+(2\pi I)^2/L^2}$. Using 
\begin{equation*}
  E_c=E_c^0-2\frac{m_c\mu}{E_c^0}  (u-\bar{u}_{ac}^a)+O(e^{-2\mu L}) 
\quad\quad \textrm{and} \quad\quad E_1E_2=\frac{m_a^2}{m_c^2}E_c^2-\mu^2
\end{equation*}
the final result is given by
\begin{eqnarray}
 \nonumber
 && F^{\mathcal{O}}_c(I,L)=
\frac{F_c^{\mathcal{O}}}{\sqrt{E_c^0L}}
-\frac{2i\Gamma_{aa}^c \left(F^{\mathcal{O}}_{aa}\right)'}{ \sqrt{E_c^0L}} (u-\bar{u}_{ac}^a)+
\\
&&
+\frac{F_c^{\mathcal{O}}}{ \sqrt{E_c^0L}}
\left[
\frac{2S_{aa}^{c,0} }{\left(\Gamma_{aa}^c \right)^2}
+\frac{m_c\mu}{(E_c^0)^2}-
\left(\frac{m_a^2}{m_c^2}E_c^0-\frac{\mu^2}{E_c^0}\right)L
 \right](u-\bar{u}_{ac}^a) +O(e^{-2\mu L})
\label{ff_mu_tag_1}
\end{eqnarray}
with
\begin{equation*}
  u-\bar{u}_{ac}^a=\pm \frac{1}{2}\left(\Gamma_{aa}^c\right)^2
  e^{-\mu_{aa}^cL \sqrt{1+\left(\frac{\pi I}{m_aL\cos(\bar{u}_{ac}^a)}\right)^2}}  
\end{equation*}

\renewcommand{\refname}{References}

\begin{figure}
% becsuletes_en_korr_1.plt
  \centering
\subfigure[$I=0$]{\includegraphics[scale=0.3,angle=-90]{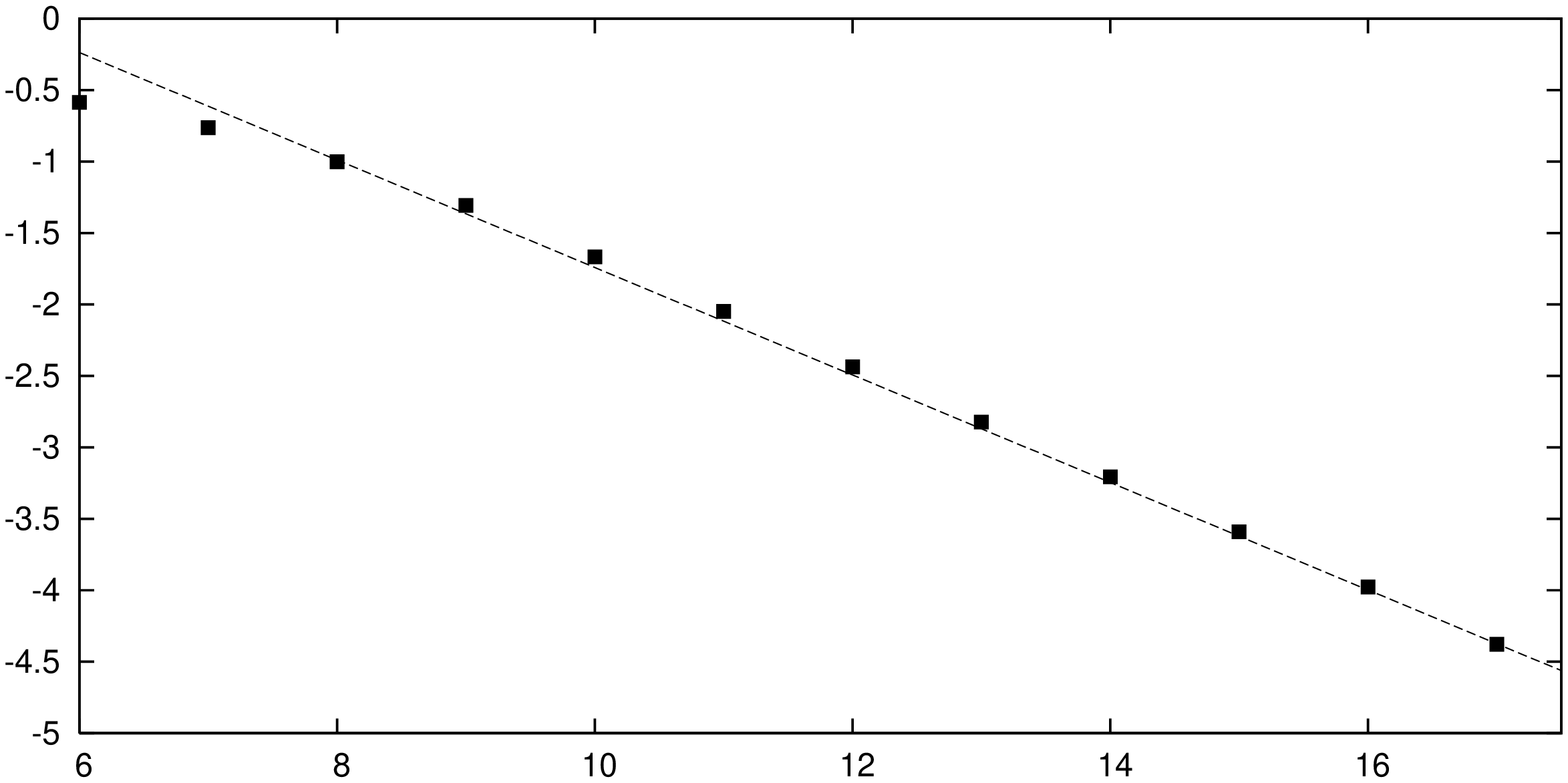}}
\subfigure[$I=1$]{ \includegraphics[scale=0.3,angle=-90]{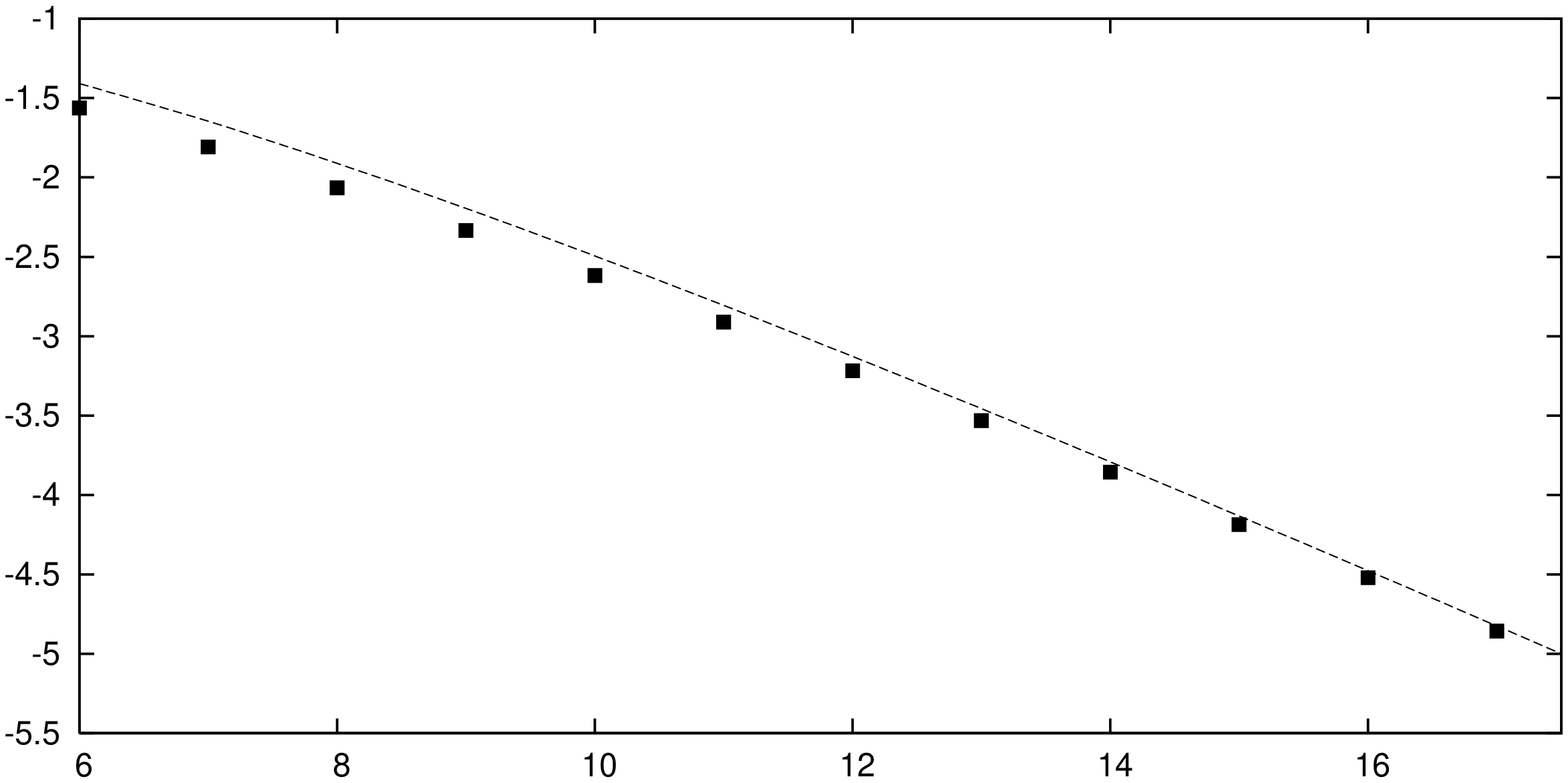}}

\subfigure[$I=2$]{ \includegraphics[scale=0.3,angle=-90]{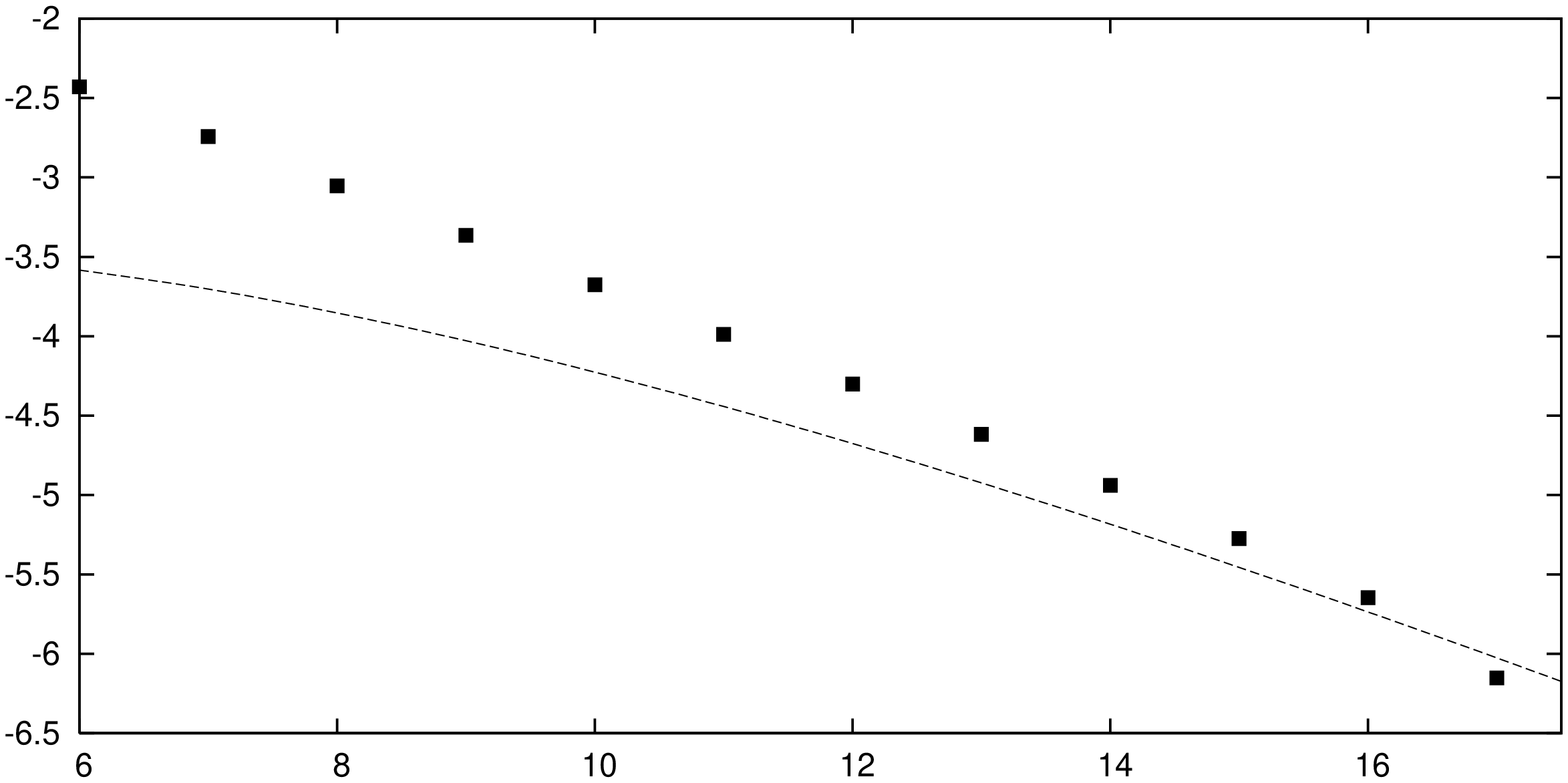}}
\subfigure[$I=3$]{\includegraphics[scale=0.3,angle=-90]{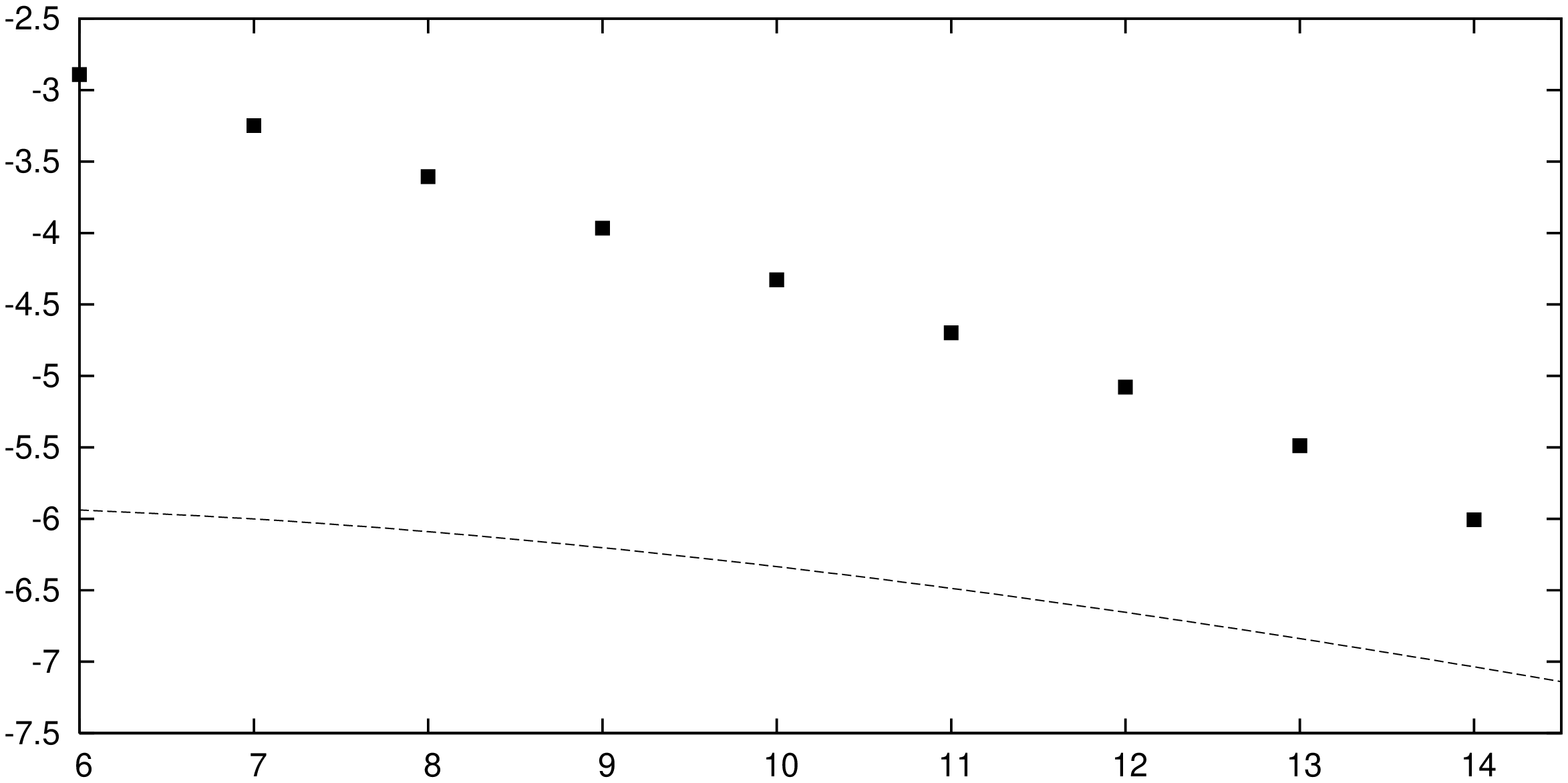}}
\caption{
Finite size corrections to $A_1$ one-particle levels in sectors
$I=0\dots 3$, $log_{10}\Delta e$ is plotted as a function of the
volume. Dots represent TCSA data, while the lines show the
$\mu$-term corresponding to the $A_1A_1\to A_1$ fusion.
\label{fig:A1_energia}}
\end{figure}

\begin{figure}
% becsuletes_en_korr_2.plt
  \centering
\subfigure[$I=0$]{  \includegraphics[scale=0.3,angle=-90]{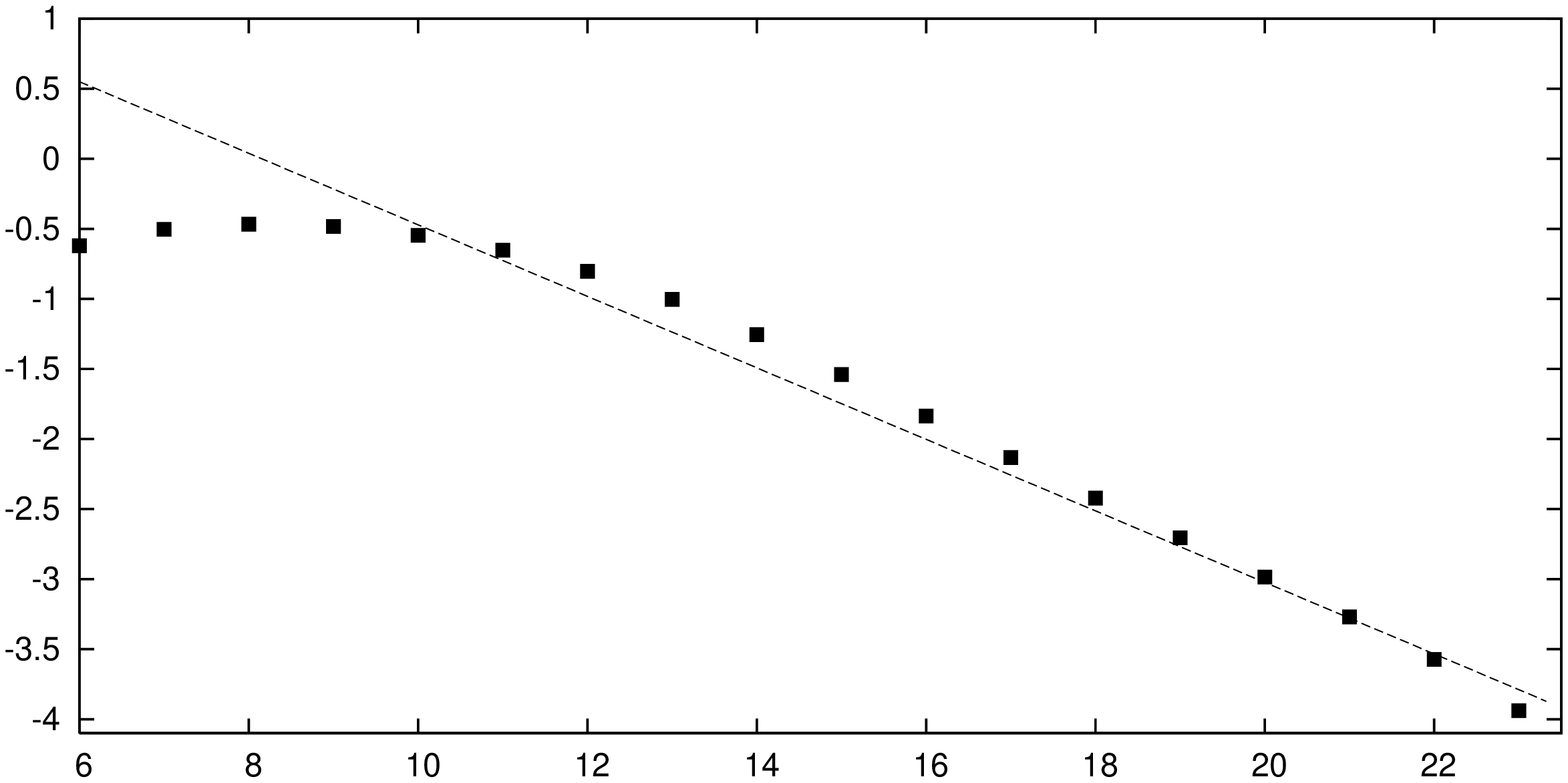}}
\subfigure[$I=1$]{ \includegraphics[scale=0.3,angle=-90]{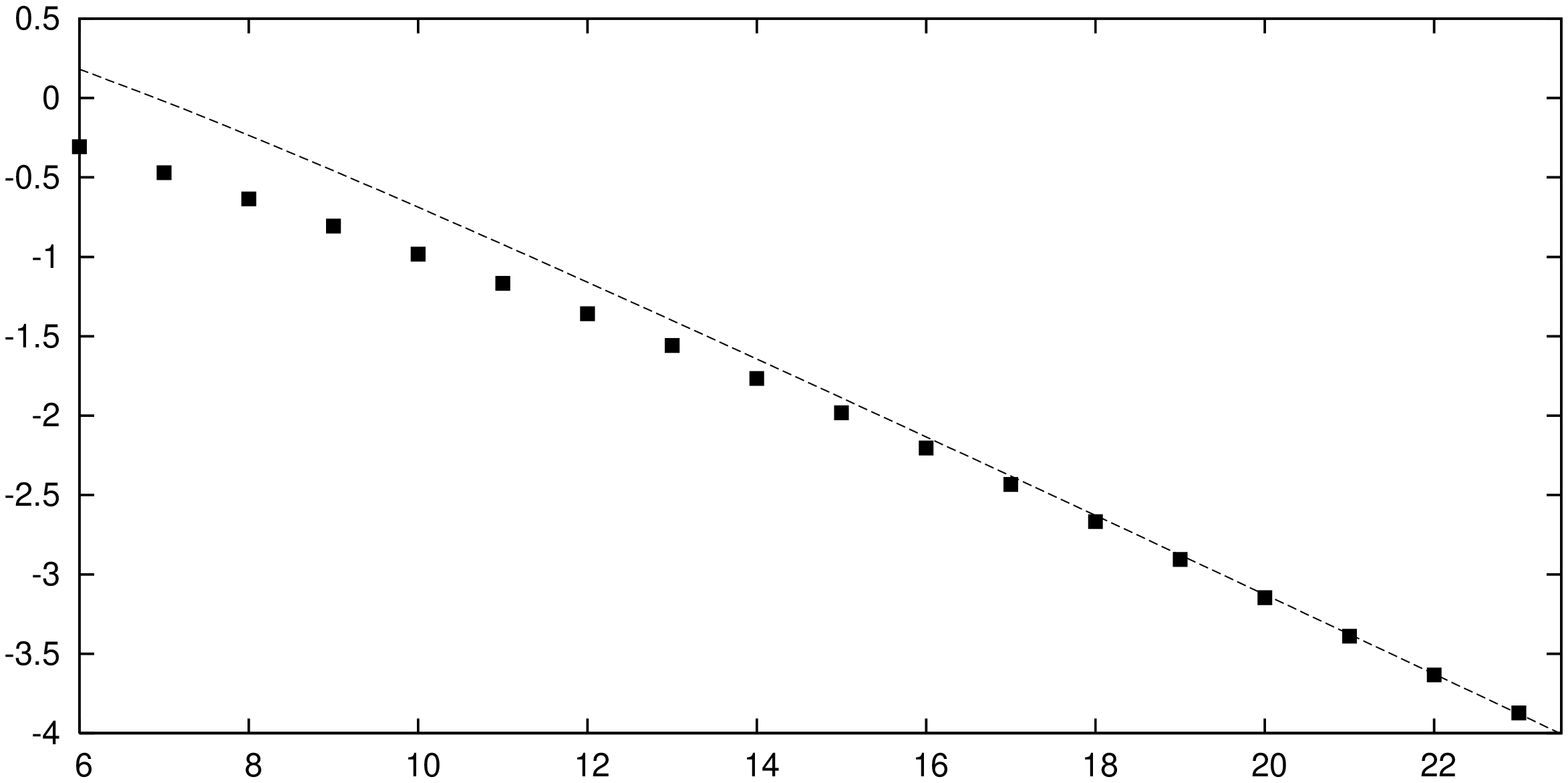}}

\subfigure[$I=2$]{ \includegraphics[scale=0.3,angle=-90]{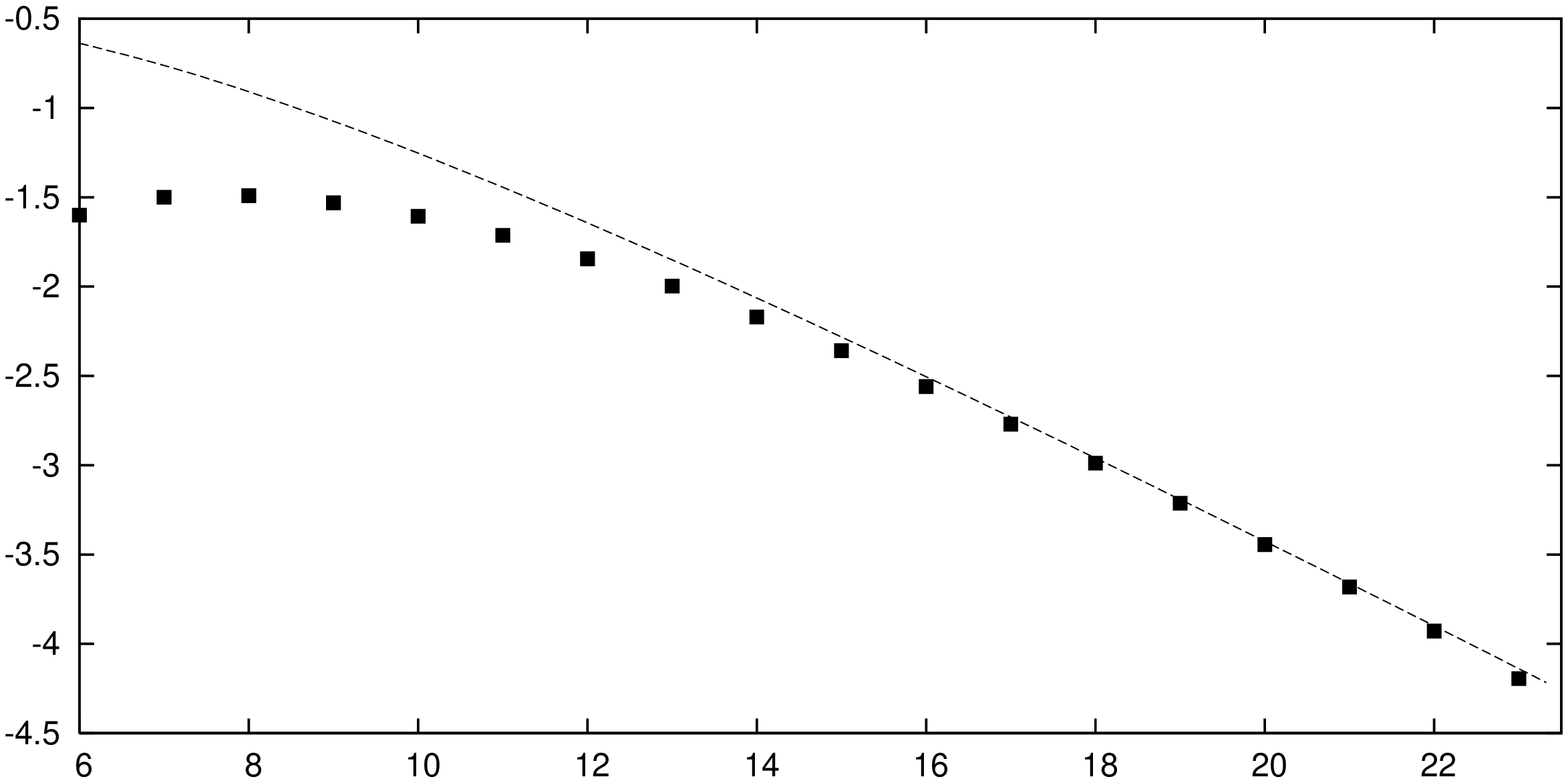}}
\subfigure[$I=3$]{\includegraphics[scale=0.3,angle=-90]{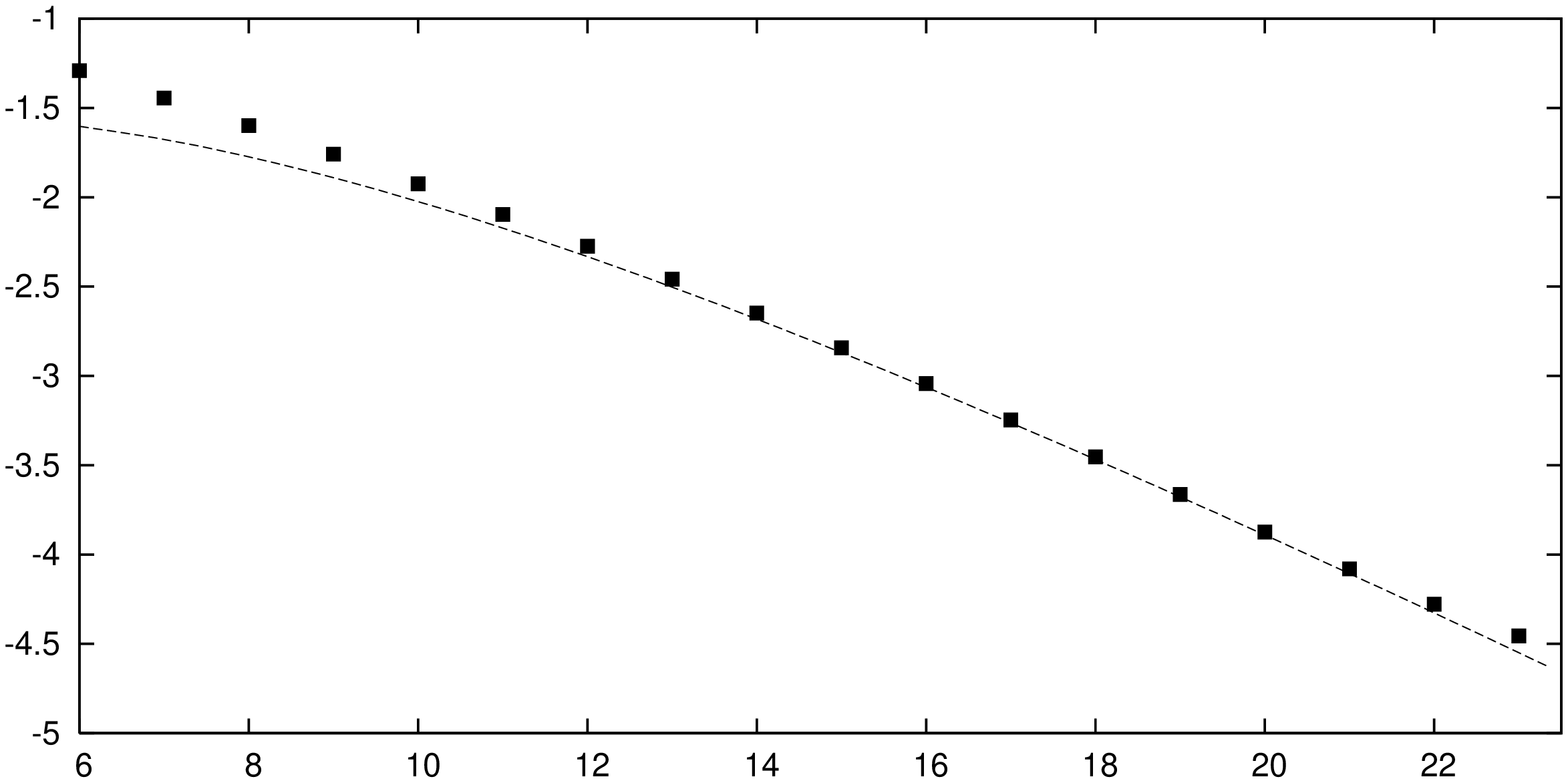}}
\caption{
Finite size corrections to $A_2$ one-particle levels in sectors
$I=0\dots 3$,  $log_{10}\Delta e$ is plotted as a function of the
volume. Dots represent TCSA data, while the lines show the
$\mu$-term corresponding to the $A_1A_1\to A_2$ fusion.
\label{fig:A2_energia}}
\end{figure}

\begin{figure}
% enkorr2_A1A3.plt
  \centering
\subfigure[$\ket{\{1,0\}}_{13}$] { \includegraphics[scale=0.3,angle=-90]{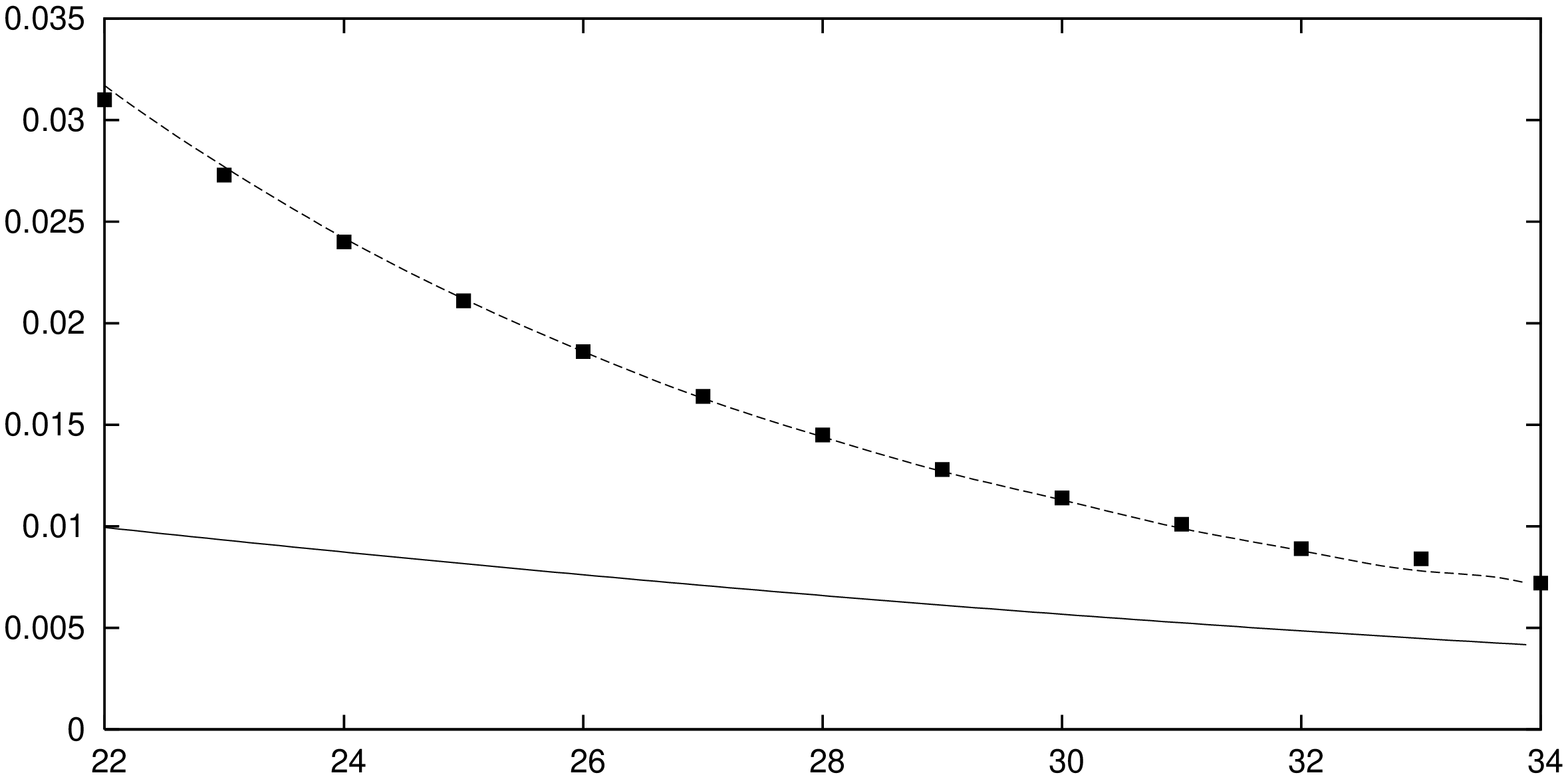}}
\subfigure[$\ket{\{0,2\}}_{13}$]{ \includegraphics[scale=0.3,angle=-90]{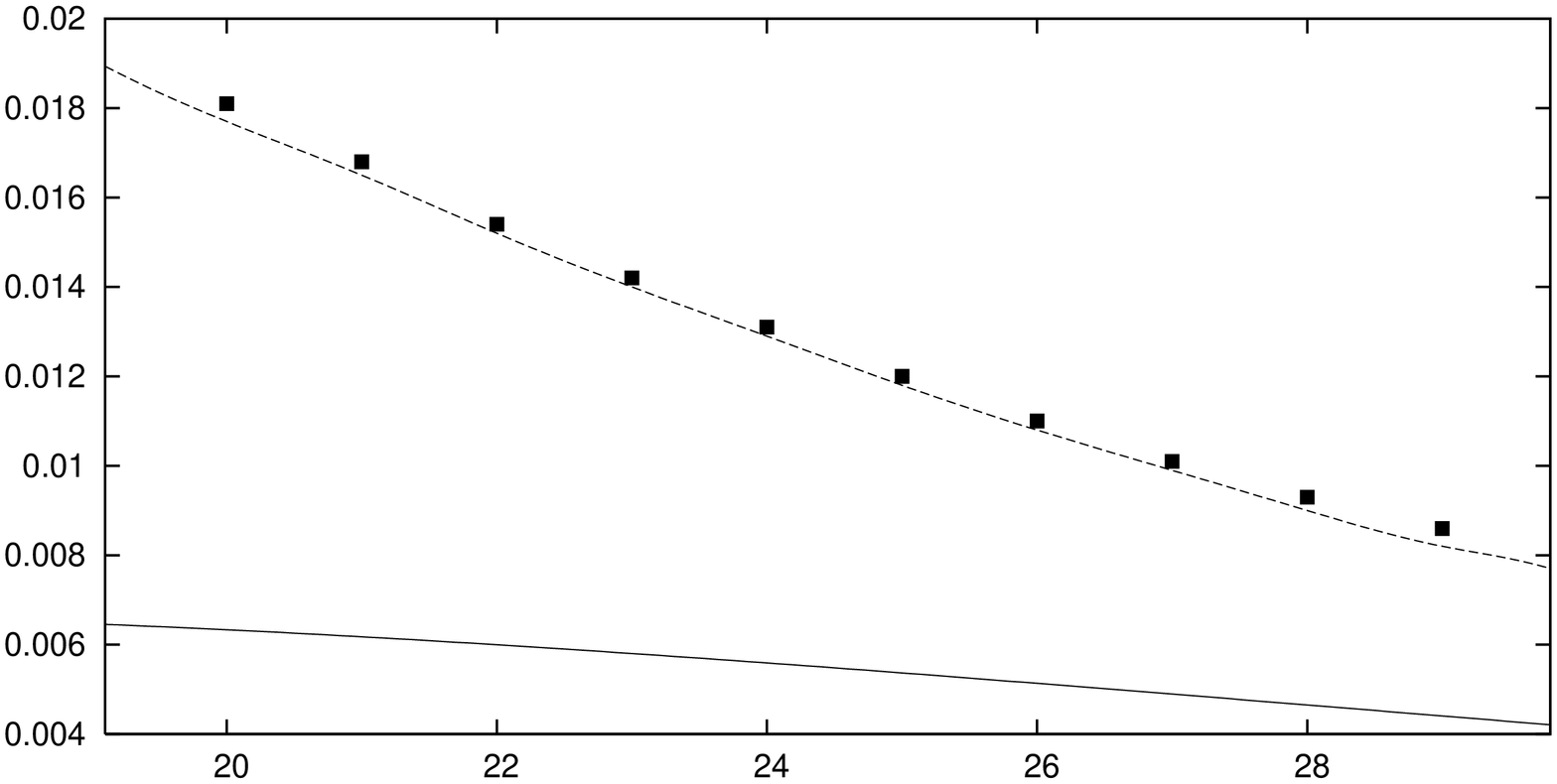}}

\subfigure[$\ket{\{0,1\}}_{13}$]{ \includegraphics[scale=0.3,angle=-90]{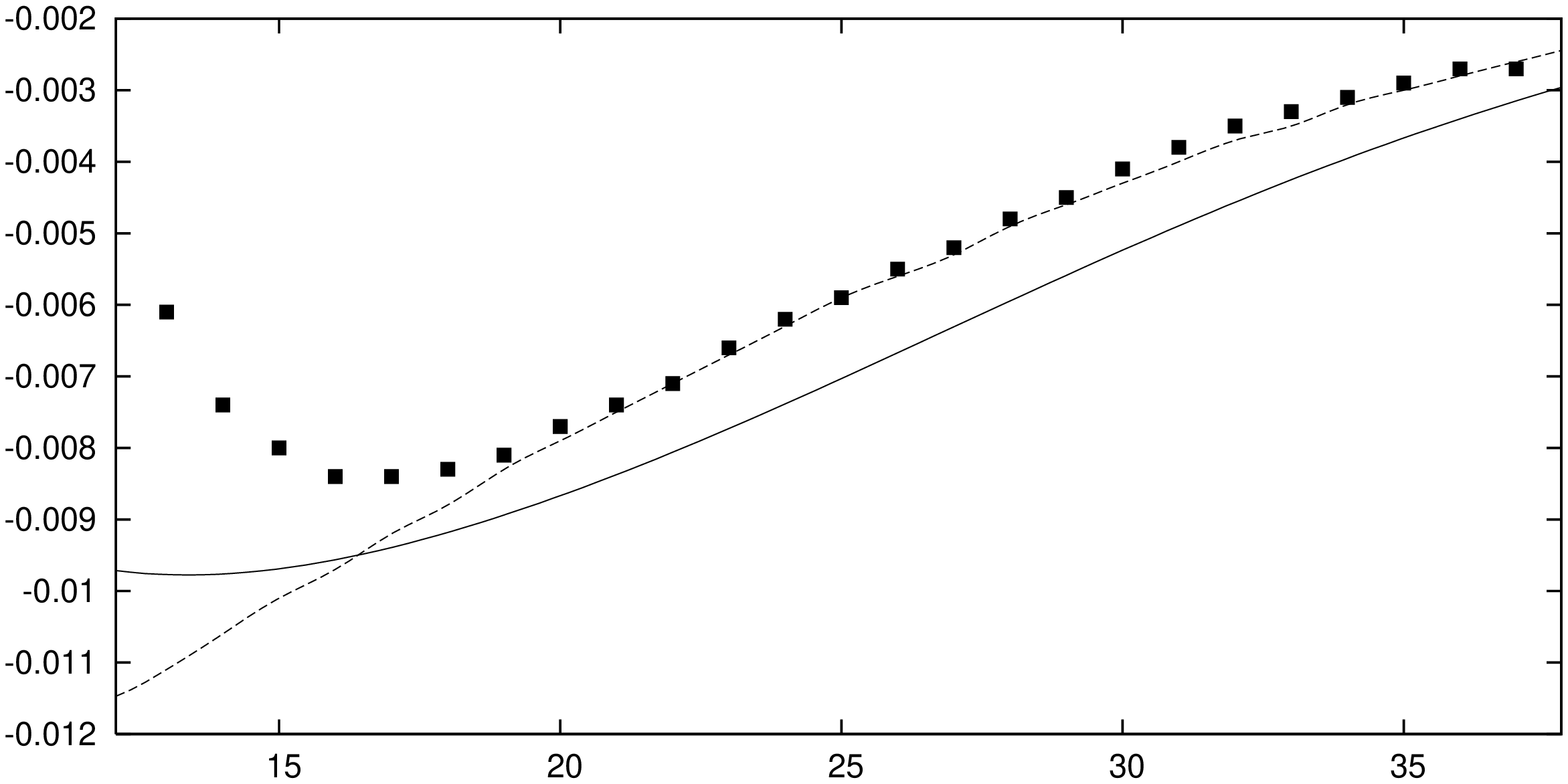}}
\subfigure[$\ket{\{1,1\}}_{13}$]{\includegraphics[scale=0.3,angle=-90]{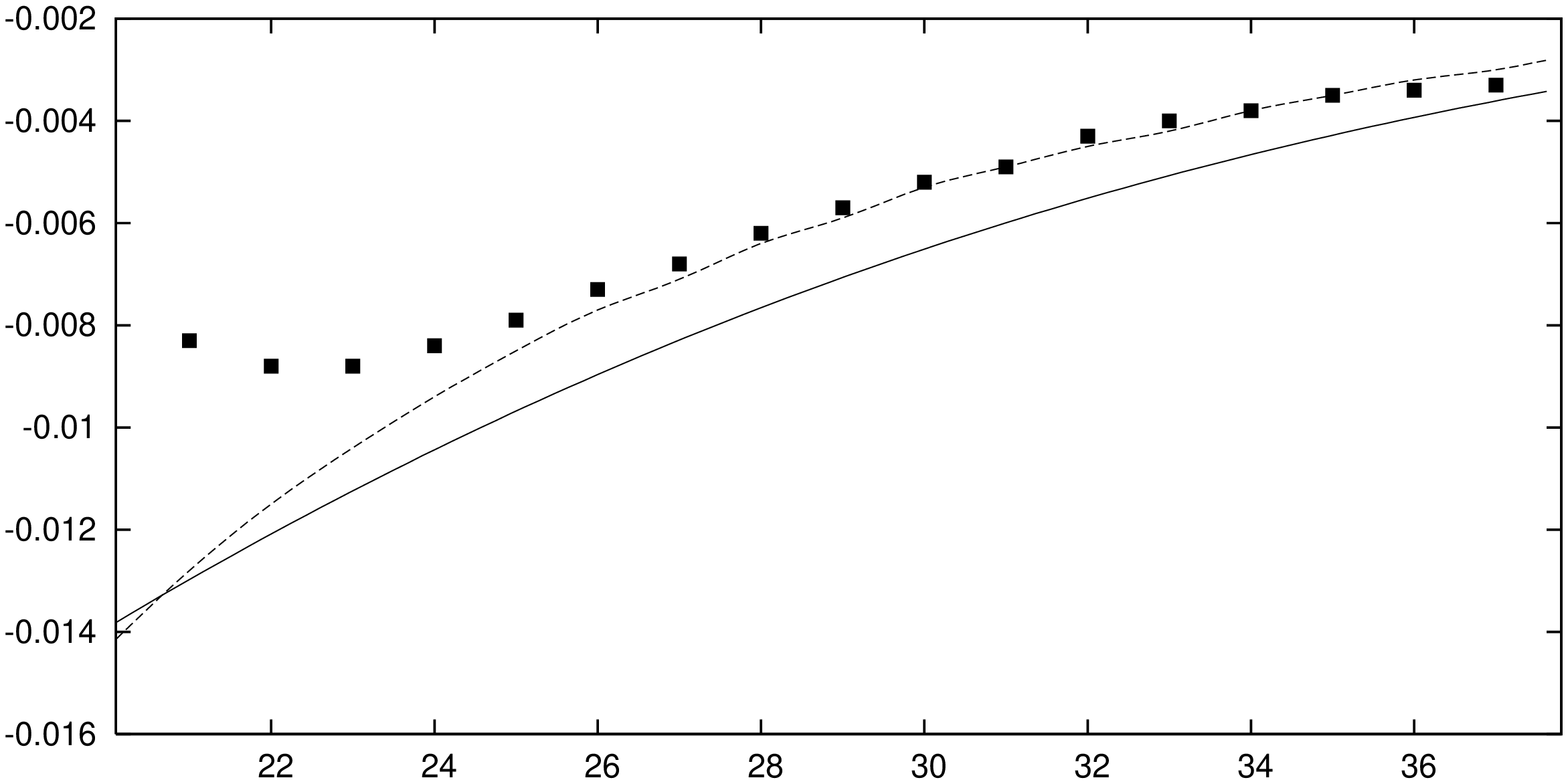}}
\caption{
Finite size corrections to $A_1A_3$ scattering states as a function of
the volume. Dots represent
TCSA data, the solid line shows the $\mu$-term corresponding to the
$A_1A_1\to A_3$ fusion. The dotted lines are obtained by the exact solution
of the quantization condition for the $A_1A_1A_1$ three-particle system.
 \label{fig:A1A3_enkorr}}
\end{figure}

\begin{figure}
% enkorr2_A1A1.plt
  \centering
\subfigure[$\ket{\{-0.5,1.5\}}_{11}$]{  \includegraphics[scale=0.3,angle=-90]{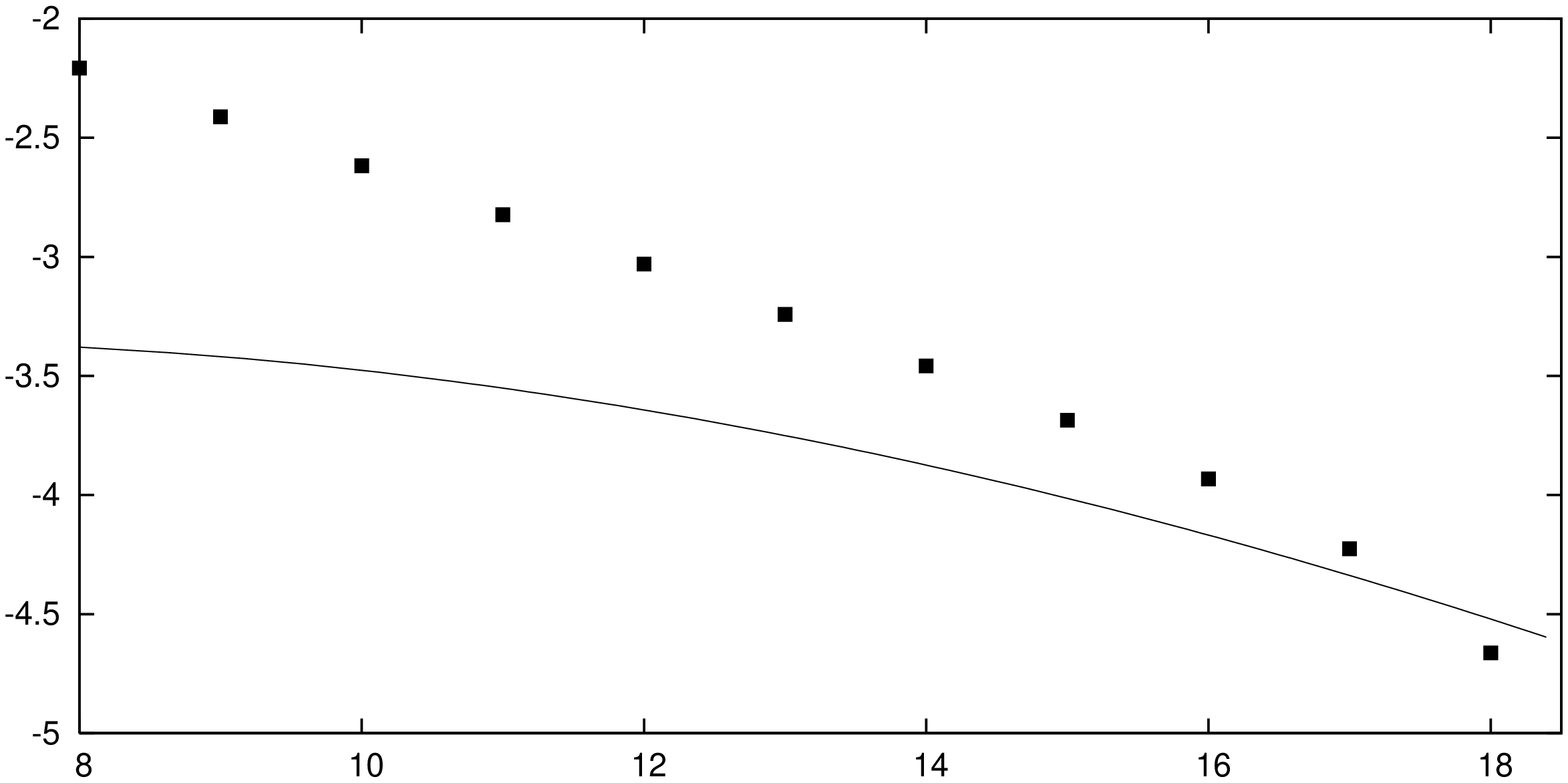}}
\subfigure[$\ket{\{-0.5,1.5\}}_{11}$]{ \includegraphics[scale=0.3,angle=-90]{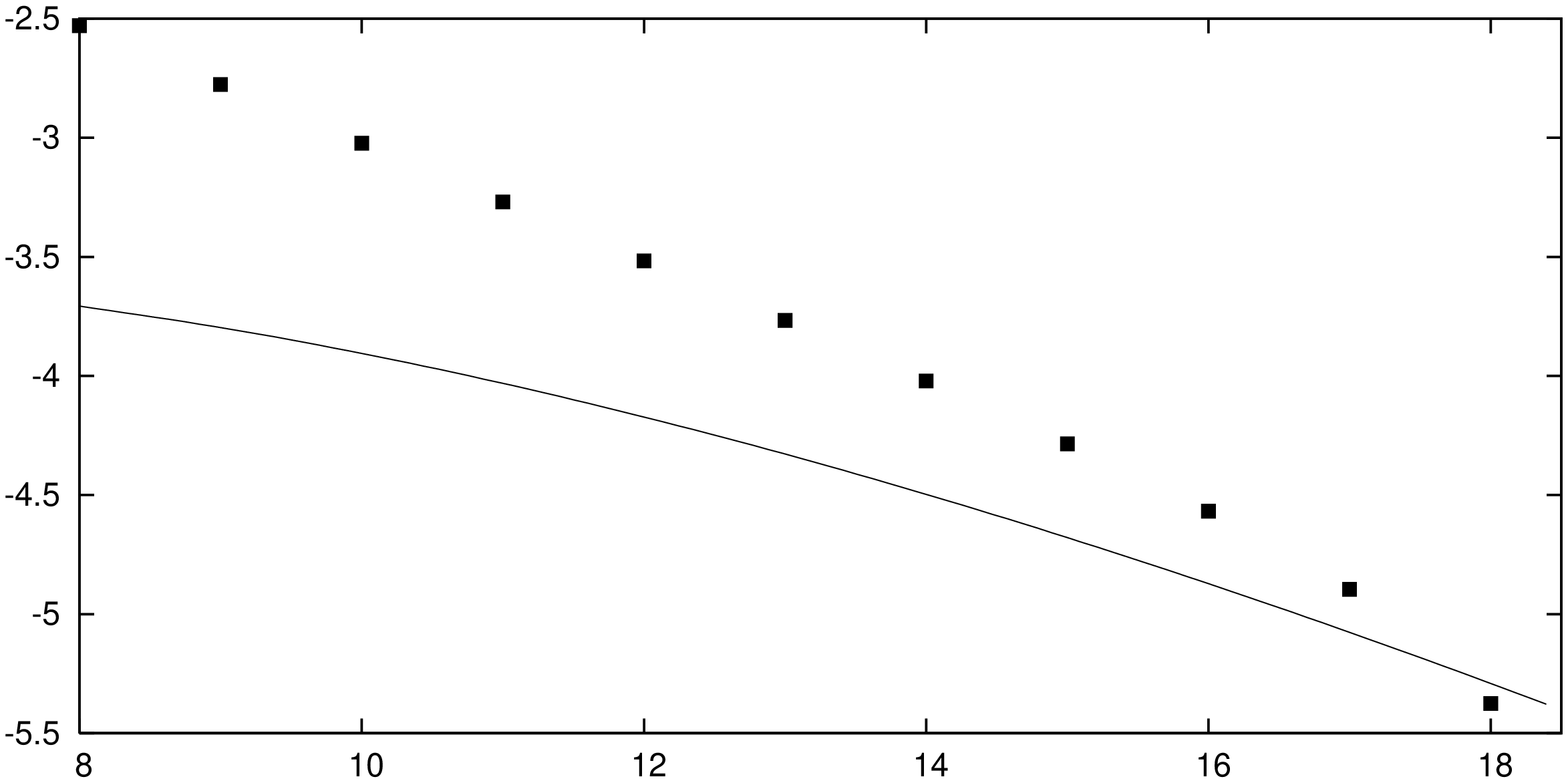}}

\subfigure[$\ket{\{0.5,1.5\}}_{11}$]{ \includegraphics[scale=0.3,angle=-90]{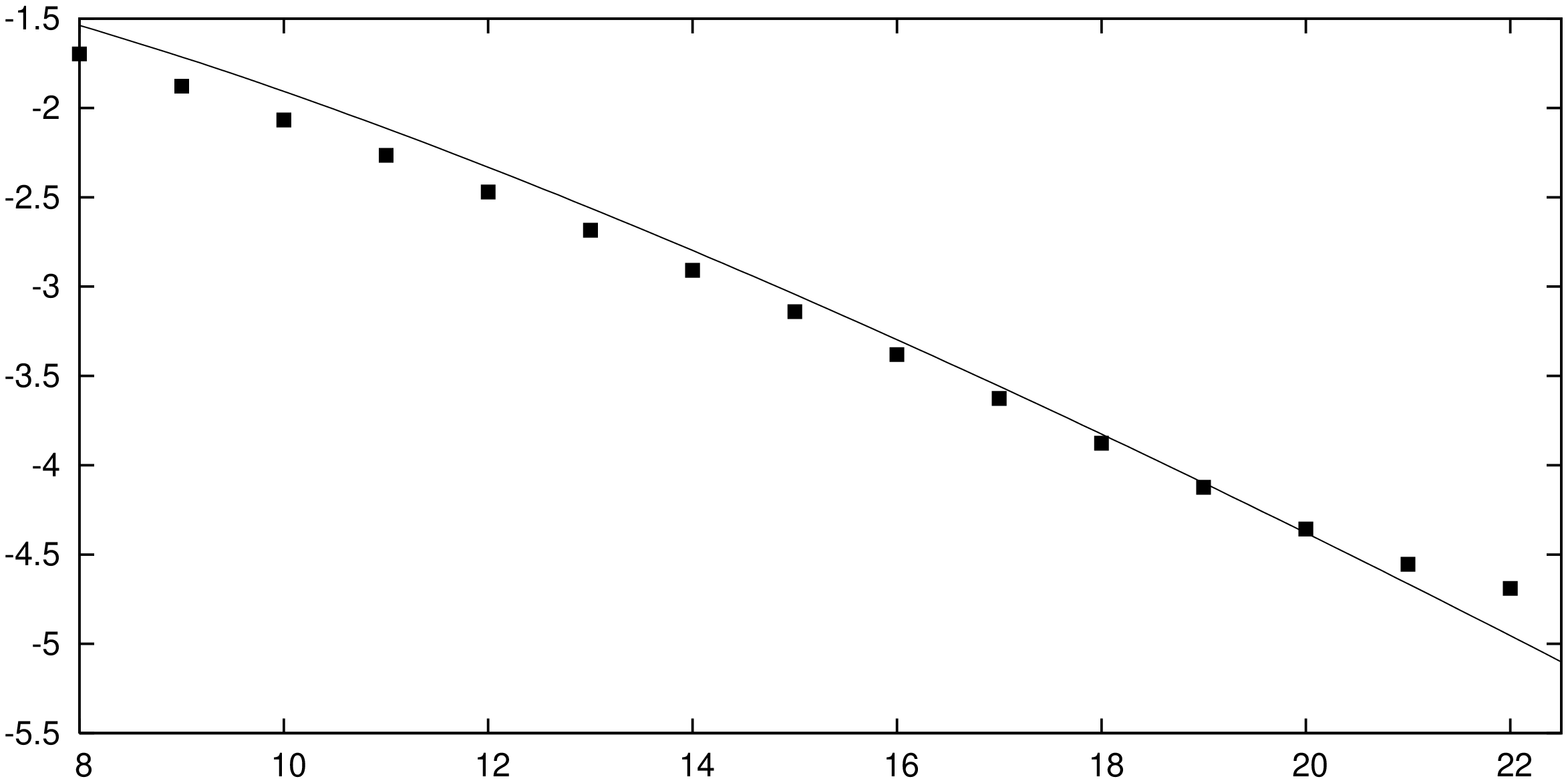}}
\subfigure[$\ket{\{0.5,2.5\}}_{11}$]{\includegraphics[scale=0.3,angle=-90]{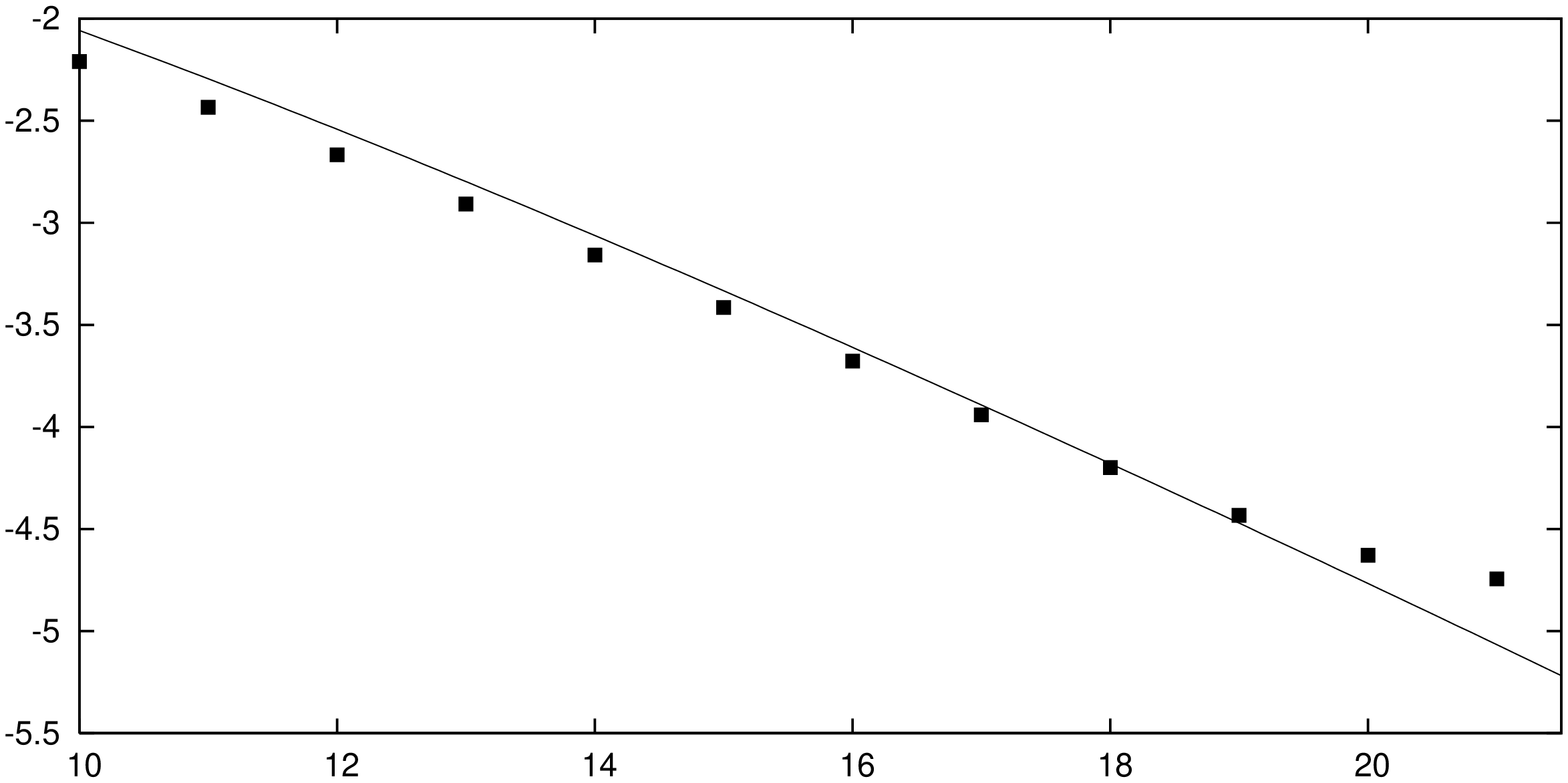}}
\caption{
Finite size corrections to $A_1A_1$ scattering states,  $log_{10}\Delta e$ is plotted as a function of the
volume. Dots represent
TCSA data, while the solid line show the sum of the two $\mu$-terms corresponding to the
$A_1A_1\to~A_1$ fusions. 
%the following is my story 
\label{fig:A1A1_enkorr}}
\end{figure}

\begin{figure}
% enkorr2_A1A2.plt
  \centering
\subfigure[$\ket{\{0,1\}}_{12}$]{  \includegraphics[scale=0.3,angle=-90]{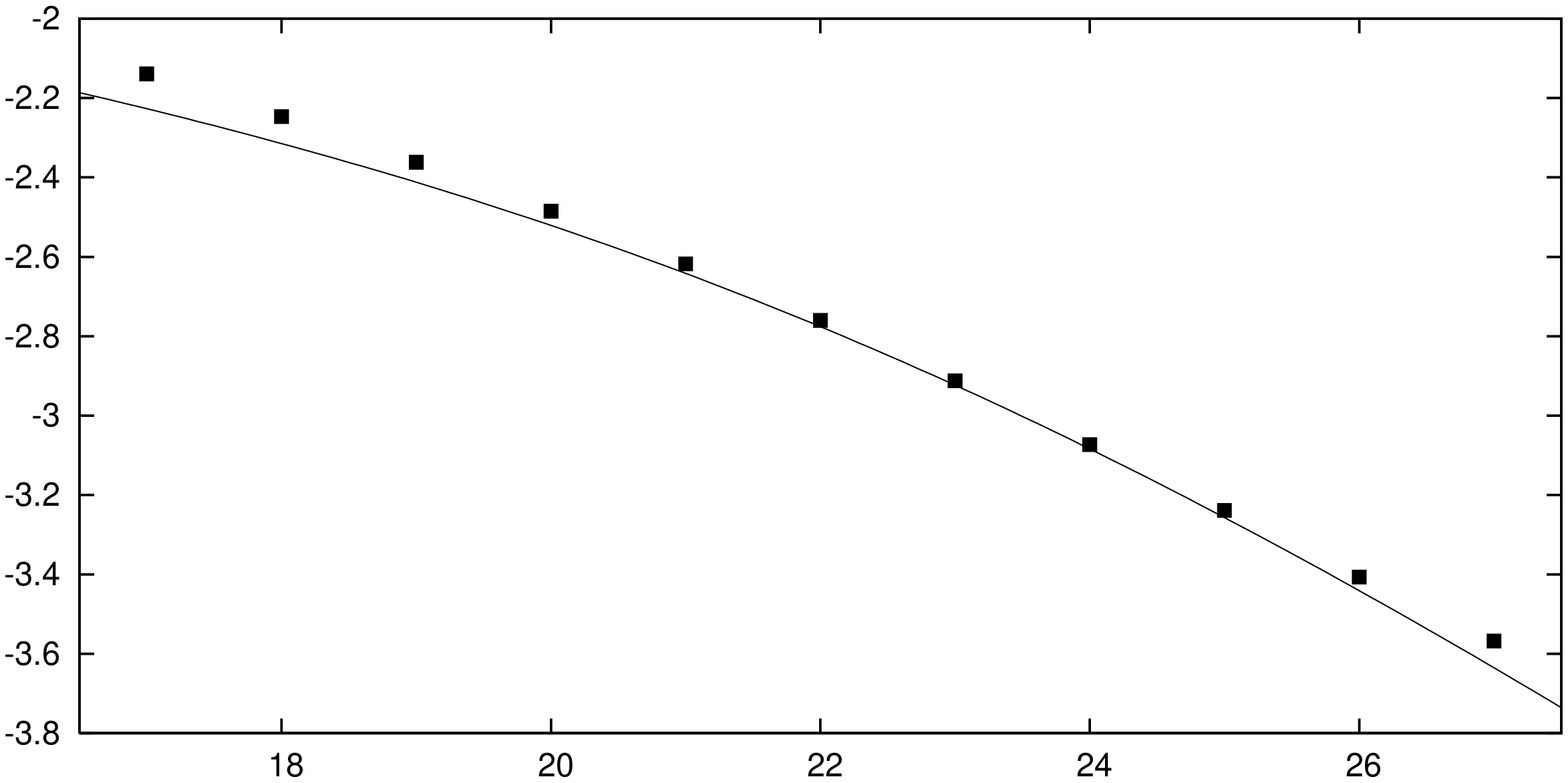}}
\subfigure[$\ket{\{1,1\}}_{12}$]{ \includegraphics[scale=0.3,angle=-90]{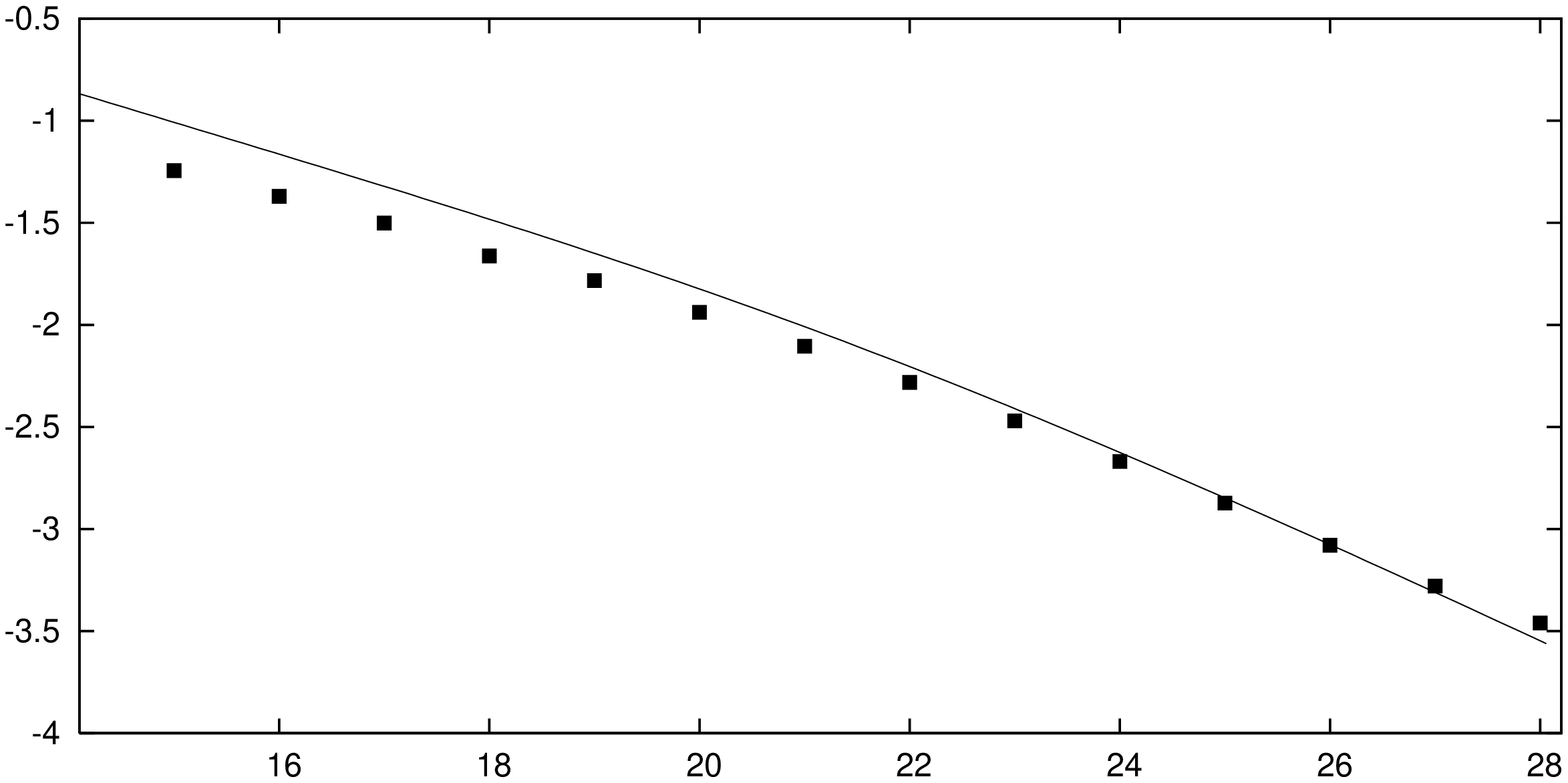}}
\caption{
Finite size corrections to $A_1A_2$ scattering states,  $log_{10}\Delta e$ is plotted as a function of the
volume. Dots represent
TCSA data, while the solid line show the $\mu$-term corresponding to the $A_1A_1\to A_2$ fusion. 
\label{fig:A1A2_enkorr}}
\end{figure}

\begin{table}
% formaz1.awk  ../../tfcsamatrixok/exponential/megoldas3_A1A1A1_s1_1.dat 
  \centering
\small
  \begin{tabular}{|c||c|c|c||c|c|}
\hline
$l$ & $\theta_1$ & $\theta_2$  & $\theta_3$ & $e(l)$ (predicted) & $e(l)$ (TCSA) \\
\hline
\hline
22 & 0.44191 & 0.68235 & -0.58742 &  3.51877 & 3.51900 \\ 
\hline 
23 & 0.42574 & 0.64280 & -0.55190 &  3.46201 & 3.46219 \\ 
\hline 
24 & 0.60523 & 0.41155 & -0.51892 &  3.41239 & 3.41255 \\ 
\hline 
25 & 0.56934 & 0.39931 & -0.48831 &  3.36890 & 3.36907 \\ 
\hline 
26 & 0.38910 & 0.53475 & -0.45989 &  3.33071 & 3.33089 \\ 
\hline 
27 & 0.50092 & 0.38119 & -0.43350 &  3.29709 & 3.29729 \\ 
\hline 
28 & 0.46686 & 0.37634 & -0.40899 &  3.26743 & 3.26767 \\ 
\hline 
29 & 0.42947 & 0.37739 & -0.38622 &  3.24122 & 3.24164 \\ 
\hline 
30 & 0.38646 $+$ 0.02332 i & 0.38646 $-$ 0.02332 i & -0.36505 &  3.21801 & 3.21832 \\ 
\hline  
31 & 0.37059 $+$ 0.04042 i & 0.37059 $-$ 0.04042 i & -0.34537 &  3.19741 & 3.19775 \\ 
\hline 
32 & 0.35572 $+$ 0.05104 i & 0.35575 $-$ 0.05104 i & -0.32706 &  3.17908 & 3.17945 \\ 
\hline 
33 & 0.34182 $+$ 0.05891 i & 0.34182 $-$ 0.05891 i & -0.31002 &  3.16275 & 3.16313 \\ 
\hline 
34 & 0.32876 $+$ 0.06513 i & 0.32876 $-$ 0.06513 i & -0.29415 &  3.14816 & 3.14856 \\ 
\hline 
35 & 0.31650 $+$ 0.07022 i & 0.31650 $-$ 0.07022 i & -0.27936 &  3.13511 & 3.13547 \\ 
\hline 
36 & 0.30498 $+$ 0.07446 i & 0.30498 $-$ 0.07446 i & -0.26556 &  3.12341 & 3.12235 \\ 
\hline 
 \hline
 \end{tabular}
  \caption{An example for the dissociation of the $A_1A_1$ bound state
  inside a scattering state. $\ket{\{2,0\}}_{31,L}$ is identified with $\ket{\{1,1,0\}}_{111,L}$
  and the corresponding Bethe-Yang equations is solved.
For $l<30$ there is a real $A_1A_1A_1$ three-particle state in the spectrum,
  whereas at $l\approx 30$ two of the rapidities become complex and
  the two-particle state $A_1A_3$ emerges.}
  \label{tab:en_A1A3_1}
\end{table}

\begin{figure}
% f1_2.plt
  \centering
\psfrag{Tspin0}{$I=0$}
\psfrag{Tspin1}{$I=1$}
\psfrag{Tspin2}{$I=2$}
\psfrag{Tspin3}{$I=3$}
\psfrag{l}{$l$}
\psfrag{f2}{}
  \includegraphics[scale=0.4,angle=-90]{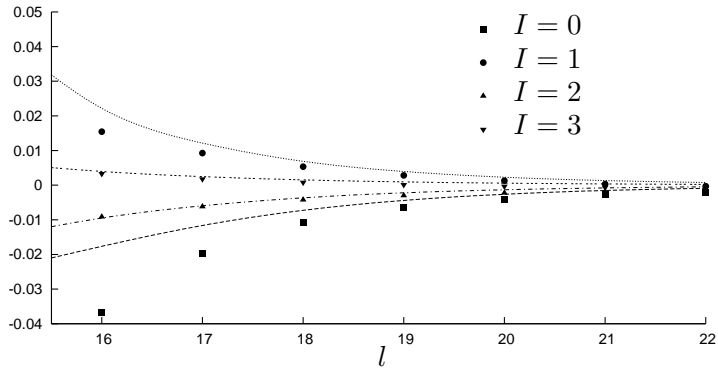}
\caption{
Finite size corrections to the elementary form factors of $A_2$. Dots
represent TCSA data, while the lines show the $\mu$-term
prediction corresponding to the $A_1A_1\to A_2$ fusion.
\label{fig:F2}}
\end{figure}

\end{document}